\documentclass{aa}  

\usepackage{graphicx}
\usepackage{txfonts}
\usepackage{bm}
\usepackage{color}

\usepackage{multicol}
\usepackage{supertabular}

\usepackage[switch]{lineno}
\graphicspath{{figures/}}

\newcommand{\Dima}[1]{{#1}}

\begin{document}

   \title{NEOForCE: Near-Earth Objects' Forecast of Collisional Events}

   %\subtitle{I. Overviewing the $\kappa$-mechanism}

   \author{D. E. Vavilov
          \inst{1,2}
          \and
          D. Hestroffer\inst{1} %\fnmsep\thanks{Just to show the usage of the elements in the author field}
          }

   \institute{LTE, Paris Observatory, univ. PSL, Sorbonne univ., univ. Lille, LNE, CNRS, 61 Av. de l'Observatoire, 75014 Paris\\
              \email{dmitrii.vavilov@obspm.fr, daniel.hestroffer@obspm.fr}
         \and
         Institute of Applied Astronomy, Russian Academy of Sciences,
              Kutuzova emb. 10, St. Petersburg, Russia\\
              \email{vavilov@iaaras.ru}
         %\and
          %  \\
         %    \email{daniel.hestroffer@obspm.fr}
             %\thanks{The university of heaven temporarily does notaccept e-mails}
             }

   \date{Received September 15, 1996; accepted March 16, 1997}

% \abstract{}{}{}{}{} 
% 5 {} token are mandatory
 
  \abstract
  % context heading (optional)
  % {} leave it empty if necessary  
   {   Robust impact monitoring of near-Earth objects is an essential task of planetary defense. Current systems such as NASA's Sentry-II, {NEODyS's} CLOMON2 and Aegis of ESA have been highly successful, but independent approaches are essential to ensure reliability and to cross-validate predictions of possible impacts, probability and path on Earth.}
  % aims heading (mandatory)
   {We present NEOForCE (Near-Earth Objects’ Forecast of Collisional Events), a new %fully 
   independent monitoring system for asteroid impact prediction. By relying on orbital solutions from DynAstVO at Paris Observatory and using an original methodology for uncertainty propagation, NEOForCE provides an alternative line of verification for future impact assessments and strengthens the overall robustness of planetary defense.}
  % methods heading (mandatory)
   {As other monitoring system NEOForCE samples several thousands of so-called 'virtual asteroids' from the uncertainty region and integrates their orbits up to 100 years into the future. 
   Instead of searching for close approaches of the virtual asteroids themselves with the Earth, our system looks for times when the Earth comes close to the ''realistic'' uncertainty regions around them, which are mostly stretched along {the osculating ellipses of virtual asteroids}. For {every} virtual asteroid {and every possible collision time} we also estimate the maximal impact probability, and only if this value is large enough {($>5\times10^{-8})$} {we} continue to the next step. In this second step, we compute how the original asteroid orbit should be slightly modified so that the new trajectory leads to an Earth impact, which allows us to confirm the possible collision and estimate the impact probability.}
  % results heading (mandatory)
   {We tested NEOForCE against NASA’s Sentry-II system on five representative asteroids with high impact probability and significant number of possible collisions: 2000~SG344, 2005~QK76, 2008~JL3, 2023~DO, and 2008~EX5. NEOForCE successfully recovered mostly all possible collisions reported by Sentry-II with impact probabilities above $10^{-7}$, demonstrating the robustness of our approach. In addition, NEOForCE identified several potential impacts at the $10^{-7}$–$10^{-6}$ level that Sentry-II did not report. }
  % conclusions heading (optional), leave it empty if necessary 
   {}

   \keywords{Celestial mechanics --
                Minor planets, asteroids: general --
                Methods: numerical
               }

   \maketitle
%
%________________________________________________________________

\section{Introduction}

Estimating the probability that a particular asteroid can hit the Earth in a foreseen future is an important task of planetary defense. The orbit of an asteroid in general does not have absolute or infinite precision, and for a newly discovered object the uncertainty of its orbital parameters can be significant. Moreover, this uncertainty increases with time if no additional observations are provided. Therefore, apart from the nominal solution of the orbital parameters (usually computed from the least-squares method), there are other solutions, which also well represent observations, sometimes with similar probability. These sets of orbital parameters are in general called virtual asteroids (VAs) \citep{2000Icar..145...12M}. Some of these VAs can collide with the Earth and some pass by safely, which {allows the impact probability to be in a range between 0 and 1}.

One of the most natural and general way to estimate the impact probability is the Monte-Carlo approach: sample a large number of VAs from the uncertainty region of orbital parameters according to the probability distribution, propagate their orbit in the future, and last derive impact probability by the fraction of colliding orbits over all considered in the sample. This approach has many advantages: It does not have additional assumptions, and also has a direct formula for estimating the uncertainty of the obtained result: $\sigma_{MC} = \sqrt{\frac{P_{MC}(1-P_{MC})}{n}}$, where $P_{MC}$ is the impact probability and $n$ is the number of considered VAs \citep{2015MNRAS.446..705V}. However, if the impact probability is small---which is generally the case with NEAs ---the number of required VAs to integrate to roughly estimate the impact probability (so that $\sigma_{MC} \lesssim 0.3 P_{MC}$) is $10/P_{MC}$. Therefore, this approach is impractical to systematic search for possible impacts with small impact probability values on a large set of of objects.
%and somehow becomes needless.

A much more effective approach is a linear approach: where one assumes that the errors of the orbital parameters at the time of a possible collision, $t$, are linearly related to the errors at the epoch of observations, $t_0$. In this case, one can integrate only the orbit of the nominal asteroid state vector together with variational equations, deriving the transition matrix and partial derivative matrix of orbital parameters at time $t$ over $t_0$. Using this matrix one can estimate the uncertainty region of the asteroid at time $t$ and assess the impact probability. The first linear method was introduced by \citet{1993BAAS...25.1236C} who also proposed to use the target plane for the impact analysis. {In addition to the linear assumption (which can be broken by a close encounter)} the important drawback of this approach is that the asteroid must come close to the Earth (enter its sphere of influence $\approx 929,000$~km), making linearization validity limited. \citet{2015MNRAS.446..705V} proposed the use of a special curvilinear coordinate system related to the asteroid's osculating orbit to overcome this limitation. This allows us to calculate satisfactorily accurate probabilities of an asteroid collision with the Earth, even if the distance between the Earth and the nominal asteroid at time $t$ exceeds 1~au. Later \citet{2020MNRAS.492.4546V} improved this approach and called it Partial Banana Mapping (PBM) method. Because the method constructs the curvilinear uncertainty region of an asteroid  then finds the part of this region, which is closest to the Earth and maps it to the target plane.
However, the linear relation between the errors of orbital parameters is a very important assumption and is limiting the usage of {linear methods} to the situation when the two-body formalism is not significantly violated.

A better compromise between linear methods and a full Monte-Carlo approach was introduced by \citet{1999Icar..137..269M} as a concept of line of variations (LOV). 
Instead of sampling the full multi-dimensional phase-space, the LOV reduces the problem to a single dimension that follows the direction where uncertainties and nonlinear effects are strongest \citep{2002aste.book...55M,2005Icar..173..362M,2005A&A...431..729M}. By distributing VAs along this line and interpolating between them, one can efficiently identify possible impacting solutions and trace distinct dynamical pathways. This approach makes it possible to detect even very small impact probabilities while keeping the computational effort manageable. Because of these advantages,  operational monitoring systems at NASA, Sentry\footnote{https://cneos.jpl.nasa.gov/sentry/},  CLOMON2\footnote{https://newton.spacedys.com/neodys2/index.php?pc=4.1} (operated by {NEODyS}) and {ESA's Aegis\footnote{https://neo.ssa.esa.int/risk-list}} were all built upon the LOV technique.
Recently, NASA introduced the Sentry-II system \citep{2021AJ....162..277R}, which samples the full confidence region rather than restricting to LOV. Potential impacts are then confirmed by adding an impact pseudo-observation and refitting the orbit to match the data. These methods require several thousands of VAs' orbits to integrate for the first search of close encounters only, but can detect possible impacts with probabilities of the order of $10^{-7}$ while being much more efficient than the Monte-Carlo one.

\citet{2025A&A...699A.158V} introduced a semi-linear method for impact probability estimation, as an extension of the linear Partial Banana Mapping method \citep{2020MNRAS.492.4546V} built in previous work. This extension (called Impactor PBM or Improved PBM: IPBM) consists of a direct search for a state vector at the epoch of observations, $t_0$, that leads to a close encounter at the time of a possible collision, $t$. This method requires several orbits to be integrated, but still much less than required by Sentry of CLOMON2 systems. The method has a moderate limitation in the way that the asteroid position's uncertainty should not exceed the whole orbit: if a VA that collides with the Earth makes different number of revolutions around the Sun than the nominal one, this collision can not be found. This naturally puts a threshold for the impact probability that can be more or less guaranteed to be found (order of $10^{-5}$). This method successfully fills the gap between restricted but very efficient linear methods and robust non-linear methods that require several thousands of orbit propagation.

In this work, we want to extend the applicability of this semi-linear method so that we can find possible collision with smaller impact probability values. We call our monitoring system \textbf{N}ear-\textbf{E}arth \textbf{O}bject's \textbf{For}ecast of \textbf{C}ollisional \textbf{E}vents (NEOForCE). The idea is to take the approach of LOV and combine it with the semi-linear PBM. We sample VAs along LOV, assign smaller uncertainty regions for each VAs and apply semi-linear PBM for each VAs. The paper is organized as follows. In Introduction we briefly explain the concept of the PBM method. In Section~\ref{sec:neoforce_algorithm} we describe step by step the algorithm of our system NEOForCE. Section~\ref{sec:target_plane} is dedicated to target plane analysis used in NEOForCE. In Section~\ref{sec:results} we discuss the practical realization of the method, the completness of the system and validation by comparing with Sentry-II. This follows by a conclusion.

\subsection{Concept of the Partial Banana Mapping method}
\label{sec:pbm_concept}

We briefly outline the concept of the Partial Banana Mapping (PBM) method, which forms the core of our system. A more detailed discussion can be found in \citet{2015MNRAS.446..705V,2020MNRAS.492.4546V}.

%We have to describe the concept of Partial Banana Mapping method, as this is the core of our system. We will give here a brief idea but encourage to read about the method \citep{2015MNRAS.446..705V,2020MNRAS.492.4546V}.

\begin{figure}
	\includegraphics[width=\columnwidth]{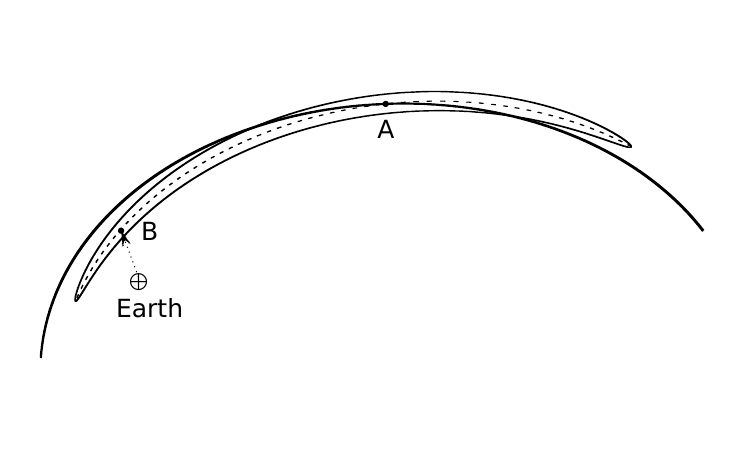}
	\caption{Schematic illustration of the confidence curvilinear ellipsoid. Point A is the nominal position of the asteroid at time $t$, and point B is the VA on the main axis of the confidence ellipsoid at the same time $t$, which is closest to the Earth after projection onto its target plane. The arrow indicates the velocity direction of Earth with respect to the confidence ellipsoid. The bold line shows the nominal orbit of the asteroid.   {Image credit: Fig.~3 in \citep{2020MNRAS.492.4546V}.}}
	\label{fig:Closest_aster}
\end{figure}

The key assumption of PBM is that, over a considered time span, the uncertainties of the orbital elements evolve almost linearly from their values at the epoch of observations $t_0$. In other words, gravitational perturbations act in a similar way on the entire uncertainty region, so that most elements change coherently. In this case, the only orbital element that strongly differs is the mean anomaly —-- the angle that determines the position of the asteroid along its orbit.
As a result, the uncertainty region is stretched primarily along the orbital path, while remaining narrow and close to the nominal trajectory. This elongated, curved region resembles the shape of a banana (see Fig.~\ref{fig:Closest_aster}).

%The Partial Banana Mapping (PBM) method is based on the linear assumption: the errors of orbital elements at future times are approximately linearly related to the ones at the time of observations, $t_0$. This means that either the asteroid is not influenced by other major bodies expept for the Sun or this influence is applied to the whole uncertainty region similar way.
%The Partial Banana Mapping (PBM) concept is based on the assumption that we have the situation when the orbital parameters for every considered part of the uncertainty region changes almost the same way. 
%In this case, the only orbital element that strongly differs is the mean anomaly —-- the angle that determines the position of the asteroid along its orbit. Because of this, the uncertainty region spreads mostly along the orbital path, becomes narrow and elongated, while still remaining close to the nominal orbit. This region is curved and looks a bit like a banana (see Fig.~\ref{fig:Closest_aster}).

To describe this ''banana-shaped'' uncertainty region more accurately, we use a special coordinate system based on the osculating elliptical elements. In this system, one of the coordinates is the mean anomaly, $M$, and the other two $(\xi, \eta)$ describe small shifts to the side and above/below the orbit \citep{2015MNRAS.446..705V}. This makes the uncertainty region easier to describe mathematically. In these curvilinear coordinates, the shape of the uncertainty region is often close to a six-dimensional Gaussian cloud.

The PBM method focuses on the part of the uncertainty region that might actually lead to an impact with the Earth. At the time of the possible impact it searches along the largest axis of the uncertainty region for the point (point B in Fig.~\ref{fig:Closest_aster}) that comes closest to the Earth, after projecting everything onto a target plane (plane perpendicular to the asteroid-Earth relative motion).
Once this closest point is found, PBM uses the local uncertainty shape around it to estimate the probability that the asteroid might hit the Earth. The method also finds the distance between point B and the Earth's center in the target plane, and the difference in $M$ coordinates for the Earth and point A (nominal position of the asteroid).

%__________________________________________________________________

\section{NEOForCE monitoring system}
\label{sec:neoforce_algorithm}

Our impact monitoring system \textbf{N}ear-\textbf{E}arth \textbf{O}bject's \textbf{For}ecast of \textbf{C}ollisional \textbf{E}vents uses a state vector (coordinates and velocities) and its {$6\times6$} covariance matrix of an asteroid at the average time of observations.
The covariance matrix $\mathbf{C}_0$ gives us an approximation of the uncertainty of the state vector  $\mathbf{x}_0$ at time $t_0$. If inside this uncertainty region there is {a possible initial state} vector that leads to a collision with the Earth, then the asteroid could potentially have an impact. The goal of a monitoring system is to find all  significantly large areas of {state vectors} that lead to collisions, \Dima{later referred as Virtual Impactors (VI)}. 

Our monitoring system consists of several stages. First, we cut the covariance matrix $\mathbf{C}_0$ by many pieces and take VAs as representatives of these pieces (Section~\ref{sec:initial_sampling} \textit{Initial sampling}). Second, for each VAs we compute all the times when the collision can happen around this VA (Section~\ref{sec:pbm_initially} \textit{Close encounter precomputation}). Third, we combine the situations from the previous step, which indicate to the same possible impact (Section~\ref{sec:selection_representatives} \textit{Selection of representatives}). Forth, we try finding a Virtual Impactor (VI) or claim that this possible collision is spurious (Section~\ref{sec:vi_search} \textit{Search for virtual impactors}). Finally, we combine similar VIs (Section~\ref{sec:combination_VIs} \textit{Combination of similar impactors}).
%We are searching for virtual impactors: coordinates and velocities that lead to an Earth impact.

\subsection{Initial sampling}
\label{sec:initial_sampling}

%How we sample VAs and how we reduce covariance matrix

In general the uncertainty region of an asteroid, especially a newly discovered one, even at the mean epoch of observations is very elongated, in one preferred direction along an eigen vector. This preferred direction is also called Line of Variations (LOV)~\citep{2002aste.book...55M}.
On the first stage we sample virtual asteroids (VAs) from the uncertainty region along the LOV. %This line is determined as the longest line in the six-dimensional uncertainty ellipsoid.
This line is determined as the eigenvector of matrix $\mathbf{C}_0$ corresponding to the largest eigenvalue. Since matrix $\mathbf{C}_0$ is symmetric and positive defined, it can be decomposed by spectral decomposition:

\begin{equation}
    \mathbf{C}_0 = \mathbf{U} \mathbf{\Lambda} \mathbf{U}^\mathrm{T}
\end{equation}

\noindent where $\mathbf{U}$ is an orthogonal matrix of the eigenvectors, and $\mathbf{\Lambda}$ is a diagonal matrix of the eigenvalues, $\mathrm{T}$ denotes the matrix transpose operation. The principal vector corresponding to the highest eigenvalue $\lambda_{11}$ determines the LOV.

We sample $n$ VAs (in general $n=5001$) on this line according to Laplace distribution so that the Laplace integral ($\frac{\sqrt{2}}{2}\exp(-\sqrt{2} |x|)$) between two subsequent points is constant (and equals the integral from the last point to infinity). We use $\sqrt{2}$ to make the dispersion of the distribution equal one. Since we consider \Dima{$n$} VAs the integral between the points equals \Dima{$b = 1/(n+1)$}. The VAs are then determined by the following:

\begin{equation}
    l_i = - \frac{1}{\sqrt{2}}\log(1 - 2i b)   \sqrt{\lambda_{11}};\ \ l_{-i} = -l_{i}; \ \ i = 0, \Dima{n/2}
    \label{eq:sampling}
\end{equation}

\noindent where $l_i$ is the difference along the LOV from the nominal state vector.

We consider each VA as a representative of its own vicinity. We assign the covariance matrix to each VA by exchanging the largest eigenvalue by $\lambda_{11}^* = \max(l_i-l_{i-1}, l_{i+1}-l_i)$. {Thus, a 1-sigma uncertainty ellipsoid around each VA would slightly touch or even include the neighbor VA.} The covariance matrix determining the uncertainty region for a VA is computed by:

\begin{equation}
    \mathrm{C}^*_0 = \mathbf{U} \mathbf{\Lambda}^* \mathbf{U}^\mathrm{T} \\
\end{equation}

\noindent where $\mathbf{\Lambda}^*$ is matrix $\mathbf{\Lambda}$ with $\lambda_{11}^*$ as the largest eigenvalue.

\subsection{Initial check}
\label{sec:initial_check}

\begin{center}
\begin{figure}[h]
	\includegraphics[width=1\columnwidth]{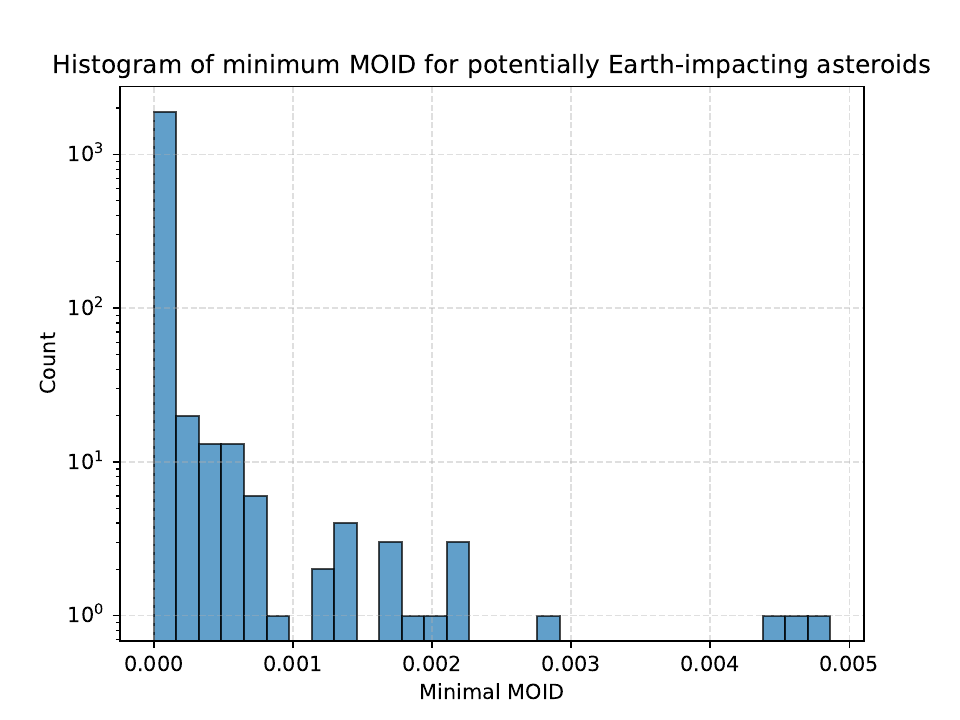} \\
    \includegraphics[width=1\columnwidth]{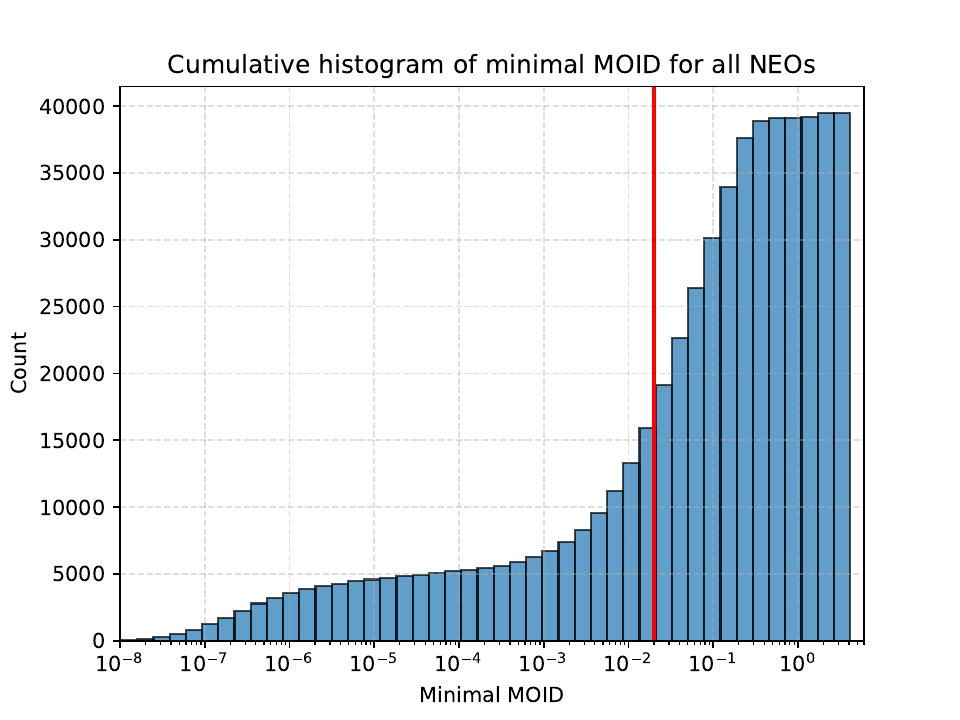}
	\caption{\textbf{Upper chart.} The histogram of minimal MOID value for asteroids presented on CNEOS web-site as ones that have non-zero impact probabilities with the Earth on September 9, 2025. \textbf{Bottom chart.} The cumulative histogram of minimal MOID for all NEOs. The vertical red line is our empirical 0.02~au threshold.}
	\label{fig:minimal_MOID}
\end{figure}
\end{center}

Instead of running NEOForCE system for all near-Earth asteroids we first fulfill the initial check whether this asteroid possibly can come close to the Earth.
We select {100} VAs along LOV (on 5-$\sigma$ interval equidistantly), integrate their orbits into the future and compute minimal orbit intersection distance (MOID) between their orbits and the Earth's orbit every {100} days. If the minimal MOID among all of them is greater than a certain threshold ({0.02}~au), we don't run the computation of the impact probability at all. {Fig.~\ref{fig:minimal_MOID} presents this minimal MOID value for all the asteroids, which are in the impact risk table on CNEOS web-site (on September 9, 2025), as well as the minimal MOID for all NEOs. As one can see our threshold 0.02~au %includes all possible Earth-impacting asteroids with large margin 
\Dima{includes all currently known NEAs with non-zero impact probability. 
At} the same time it allows us to limit the number of launches of the monitoring system to approximately 15,000 instead of 40,000. This initial check requires \Dima{less than} one minute of computations \Dima{for one asteroid} on one thread of a processor like Intel(R) Xeon(R) Gold 6230 2.10GHz.}

\subsection{Close encounter precomputation}
\label{sec:pbm_initially}

On this step for each VA we compute the information about possible close encounters. We integrate the orbit of a VA with variational equations towards the future until the final time (in general, 100 years). On each integration step we compute the osculating orbit of a VA and then obtain the minimal distance between the position of the Earth {at this time} and {the points of} this osculating ellipse (or parabola/hyperbola) numerically. We then find the local minima in these distances  indicating the time of the local minima $t$, the state vector of the VA, $\mathbf{x}^*_t$, and the Earth and the partial derivative matrix of VA's state vector at $t$ over original time $t_0$, $\mathbf{\Phi}(t_0,t)$. The covariance matrix of the VA for time $t$ is:

\begin{equation}
    \mathbf{C}^*_t = \mathbf{\Phi}(t_0,t) \ \mathbf{C}^*_0 \ \mathbf{\Phi^\mathrm{T}}(t_0,t).
\end{equation}

If the minimal distance between the osculating orbit and  the Earth is $< 0.05$~au then we assume that a collision can not be ruled out at this time $t$ for a vicinity of this VA.

%We then use the concept of the partial banana mapping method (see Sec.~\ref{sec:pbm_concept}) to find the closest point (point B) in the banana-shaped uncertainty region to the Earth. While doing so we also restrict the uncertainty region to $5\sigma$ for the search for minimal distance.

We then use the concept of the partial banana mapping method (see Sec.~\ref{sec:pbm_concept}). We construct the banana-shaped uncertainty region at time $t$ for this VA using covariance matrix $\mathbf{C}^*_t$ restricting it to 5-$\sigma$ uncertainty. In this region we find the closest point to the Earth (point B in Fig.~\ref{fig:Closest_aster}), its coordinates in $(\xi, \eta, M)$ coordinate system and compute the distance between the Earth and the projection of point B to its target plane. We also compute and save the difference in mean anomaly, $M$ coordinate, between the found point B and point A.

We also estimate the \textbf{maximal impact probability} for this VA assuming that its orbit leads to a collision with the Earth at  time $t$, approximately. For this, we compute the full covariance matrix at time $t$ around this VA:

\begin{equation}
\mathbf{\hat{C}}_t = \mathbf{\Phi}(t_0,t) \ \mathbf{C}_0 \ \mathbf{\Phi^\mathrm{T}}(t_0,t).
\label{eq:c0}
\end{equation}

Then we hypothetically put the Earth at the center of the position of the VA at time $t$ (point A in Fig.~\ref{fig:Closest_aster}) and compute the probability by  the target plane method \citep{1993BAAS...25.1236C}. To account for the distance from the VA to the nominal orbit we then multiply this value by $\exp(-\frac{1}{2} \frac{l_i^2}{\lambda_{11}})$, where $l_i$ is from Eq.~\ref{eq:sampling}. %the deviation in terms of LOV of this $i$-th VA from zero (i.e. the distance on LOV of the considered VA from the nominal asteroid). 
This probability tells us what is the maximal possible value of the impact probability with the Earth selected asteroid can have at approximately time $t$ around the vicinity of this VA. If it is lower than our threshold (nominally $5\times 10^{-8}$) we don't need to go to next stage to find virtual impactors. We consider this probability too low.

\subsection{Selection of representatives}
\label{sec:selection_representatives}

%Several closed VAs can indicate to the same collision at approximately the same time. The higher the impact probability the more VAs correspond to the same collision. Therefore on this stage we divide VAs and times computed in the previous stage into groups and select distinctive representatives.

The information obtained in the previous step often indicates that several neighboring virtual asteroids (VAs) along the Line of Variation (LOV) may correspond to the same potential encounter with the Earth. In particular, if a given close-approach time $t$ is found, multiple VAs clustered around each other can all signal the possibility of a collision at that epoch. To avoid redundancy and to efficiently characterize each event, we identify a small set of representative VAs.

\begin{figure}[h]
	\includegraphics[width=\columnwidth]{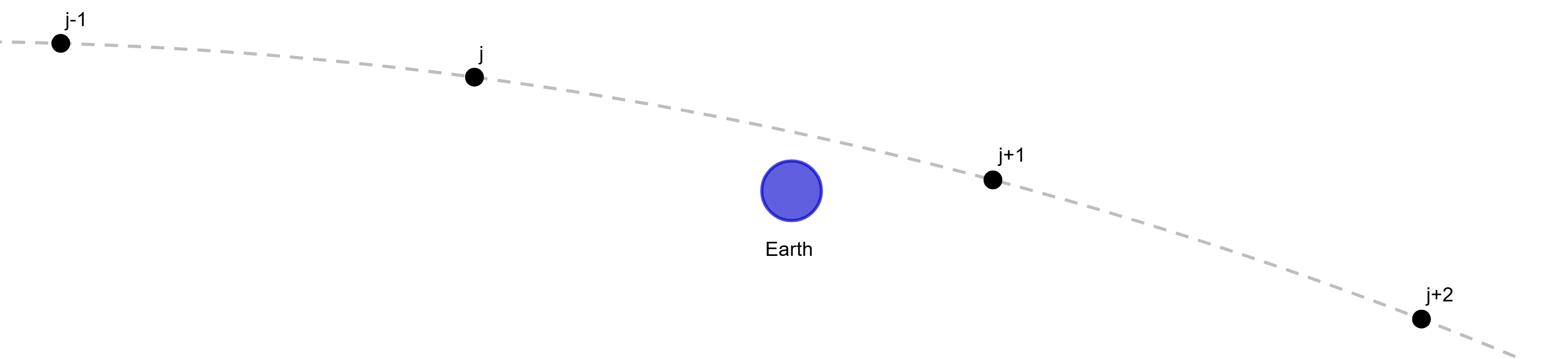} \\
    \includegraphics[width=\columnwidth]{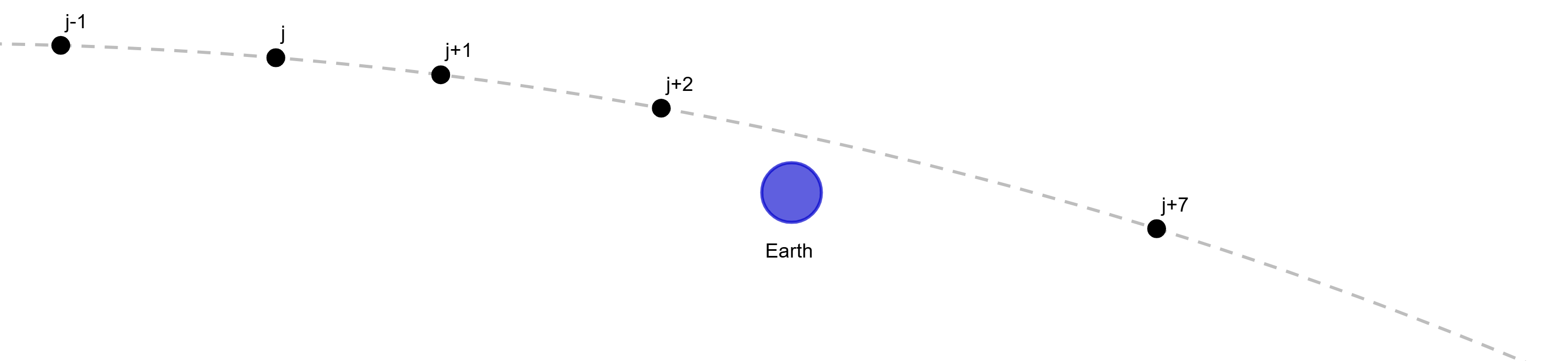}
	\caption{Schematic illustration of the different scenarios for grouping of VAs. The top chart shows the situation when the Earth is between two subsequent VAs (difference in $M$ has opposite signs). In this case we choose j+1-th VA as our primary representative and j-th as a secondary one. The bottom chart shows the situation when all the subsequent VAs are on one side from the Earth (difference in $M$ are all positive). In this situation we choose j+2-th as our primary representative. Note, that j+7-th VA is in another group but it will also be a representative of its group.}
	\label{fig:grouping}
\end{figure}

%If several subsequent VAs indicate for a possible collision at approximately time $t$ we have two options. Either VAs locate at different

For each candidate epoch $t$, we first collect all VAs that point to a possible collision at approximately that time withing 10 days. We then divide them by groups so that each group consists of subsequent VAs along LOV. Within such a group we search for two scenarios:

\begin{itemize}
    \item \textbf{Earth located between two consecutive VAs} (upper chart in Fig.~\ref{fig:grouping}). \\
    If the difference in mean anomaly, $M$, computed in Section~\ref{sec:selection_representatives}, shows opposite signs for two consecutive VAs, this implies that the Earth lies between them at the given time $t$. In this case, we select the VA that is closest to the Earth at approximately time $t$. This VA becomes the selected representative. Note that this situation can happen several times within one group if the spread of VAs exceeds several revolutions around the Sun. The second VA we consider as 'sidekick'.
    \item \textbf{Earth located on one side of the group} (bottom chart in Fig.~\ref{fig:grouping}). \\
    If the Earth is not bracketed by two VAs (i.e., all $M$ differences have the same sign within the group), then we select only one VA --- the one lying on the outer edge of the group and at the minimal distance to the Earth. This VA becomes the representative for the entire group.
\end{itemize}

Once we selected the representatives we apply additional filters to them. First, as was mentioned above, we consider only the ones with maximal impact probability $> 5\times 10^{-8}$. Second, the distance between the Earth and the curved uncertainty region (distance between point B and the Earth) for at least one of the VAs must be smaller than $0.002$~au. In the case where the Earth is located between two subsequent VAs even though we select only the closest to the Earth, %we required these two conditions to be, however, 
the representative is still selected if these two conditions are fulfilled for one of the VAs.
If the conditions are not met, we do not expect that the collision can happen or that the probability is significant. And third, we only consider the ones for which the distance between the Earth and the VA at the time of the possible collision $t$ smaller than $1$~astronomical unit. This limit with 1~au distance allows us to exclude the situations where the signs of $M$ coordinate differences are opposite but the actual positions of the VAs are extremely far from the Earth (on the different side of the osculating orbit).

This filter allows us to consider only the situation where we believe the collision can happen, and only significant ones with sufficient impact probabilities. 
After this preselection, the next stage will find explicit orbital parameters around each selected VA that can lead to a collision, and estimate the impact probability.

\subsection{Search for virtual impactors}
\label{sec:vi_search}

Again, we follow the idea of \citet{2005Icar..173..362M} in CLOMON2 system, that in order to confirm a possible collision, one must explicitly find a virtual impactor (a state vector at initial epoch, $t_0$, which actually leads to a collision). Therefore, in this stage  we try to find the virtual impactor (VI) around each representative found in the previous step. 
For this we use a semi-linear approach extension of the PBM, briefly described here, for details see \citet{2025A&A...699A.158V}. 
%This is an extension of the Partial Banana Mapping method. For a detailed understanding we encourage the reader to study this article themselves, but here we briefly describe the concept.

The objective is to find the initial condition at epoch $t_0$ of the asteroid, bringing it at time $t$ to the closest point B in Fig.~\ref{fig:Closest_aster}. The linear approximation reads as follows:

\begin{equation}
\bm{\xi}_* - \bm{\xi} = \left[ \frac{\partial \bm{\xi}}{\partial \bm{x_0}} \right] ( \bm{x}_{0*}-\bm{x}_0 ),
\label{eq:find_VA}
\end{equation}

\noindent where $\bm{\xi}$ and $\bm{\xi}_*$ are vectors in $(\xi, \eta, M)$ coordinate system at time $t$ of the asteroid (point A) and VA closest to the Earth on the main axis of curvilinear uncertainty ellipsoid (point B), correspondingly. $\bm{x}_0$ and $\bm{x}_{0*}$ are the state vectors of the selected VA and the one we are looking for, correspondingly, at $t_0$ and  $\left[ \frac{\partial \xi}{\partial x_0} \right]$ is the matrix of the partial derivatives of curvilinear coordinates at time $t$ over the Cartesian coordinates at $t_0$.

$( \bm{x}_{0*}-\bm{x}_0 )$ is the vector we are looking for: how to change the state vector of the selected VA on the previous stage so that after orbit integration it goes to point B at time $t$. In practice, for NEOForCE system we also {use these two additional state vectors}:
$\frac{3}{4}( \bm{x}_{0*}-\bm{x}_0 ) + \bm{x}_0$ and 
$\frac{1}{2}( \bm{x}_{0*}-\bm{x}_0 ) + \bm{x}_0$. {These state vectors are closer to $\bm{x}_0$ than $\bm{x}_{0*}$.} 
We integrate these orbits until time $t$ plus 0.5 days. If none of them collide with the Earth at approximately time $t$ we select the closest to the Earth. 
This newly selected VA is closer to the Earth at approximately time $t$ than the original one, so we  repeat the procedure using this VA as the new reference (starting from computing $\bm{\xi}$, $\bm{\xi}_*$ vectors and matrix $\left[ \frac{\partial \xi}{\partial x_0} \right]$). In this way, the method works as an iterative refinement process: at each step the candidate VA is moved closer to the Earth at $\approx t$ until one of {four} conditions is met: 
\begin{enumerate}
    \item[(i)] \Dima{a collision is found: we compute the probability by the target plane method (see Section~\ref{sec:target_plane}) ; }
    \item[(ii)] \Dima{ the solution drifts outside the 9-$\sigma$ uncertainty ellipsoid: we stop the procedure }
    \item[(iii)] \Dima{the maximum number of 10 iterations is reached: we stop the procedure; }
    \item[(iv)] \Dima{the asteroid enters the sphere of influence with the Earth ---  we run the target plane analysis (see Section~\ref{sec:target_plane}) to search for a VI. }
\end{enumerate}

{By taking two additional state vectors, which are closer to original one, instead of simply using $\bm{x}_{0*}$ we slightly mitigate some convergence problems that can happen after multiple close encounters.} In an extreme case where the stretching of the uncertainty region is very high and mismatches the linear estimation, the closest VA between these 3 cases can be further away {from the Earth} than the original one. In this case we select $\frac{1}{6}( \bm{x}_{0*}-\bm{x}_0)  + \bm{x}_0$ VA and proceed again.
%Once the VA entered the sphere of influence (approximately 929,000~km) we launch the target plane analysis (see Section~\ref{sec:target_plane}).

%Now this VA is closer to the Earth than the original one, we improve the collisional time $t$, and repeat the same process with this VA (. We continue it until we either find a possible collision or the vector becomes too far from the original one (escapes 9-$\sigma$ uncertainty ellipsoid) or reaches the maximal iteration of 15. 

If virtual impactor is found, $\bm{x}_{0*}$, we estimate its impact probability by the target plane method and multiply this value by

\begin{equation}
    \exp{\left[ -\frac{1}{2} [\bm{x}_{0*} - \bm{x_0}]^{\mathrm{T}} \mathbf{C}_0^{-1} [\bm{x}_{0*} - \bm{x_0}] \right] } , 
\end{equation}

\noindent   which takes into account that the found $\bm{x}_{0*}$ is not the nominal one. If a VI is not found we can use a 'sidekick' try to find a VI around it. We don't do it by default in order to save the performance but we keep this as an option for special cases.

\begin{figure}[h]
	\includegraphics[width=\columnwidth]{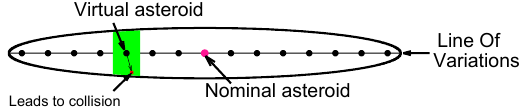} \\
    \includegraphics[width=\columnwidth]{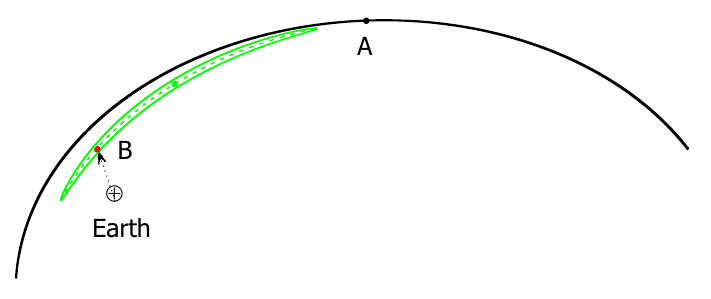}
	\caption{Schematic illustration of the work of NEOForCE system. On the upper chart one can see the nominal asteroid (purple dot), the original uncertainty region at the epoch of observation (black ellipse), the LOV (horizontal line), VAs sampled along LOV (black dots). The green area represents the smaller uncertainty region assigned to this VA. The red denotes the VI. On the bottom chart one can see the nominal position of the asteroid at time {slightly before} a possible collision, $t$ (point A), the osculating orbit of the nominal asteroid (black part of an ellipse), the Earth, the position of the VA (green dot), the uncertainty region assigned for this VA at time {slightly before} $t$ (green curved ellipse). The red dot (point B) represents the position of the VI. The dashed vector shows the relative motion of the Earth and the green uncertainty region. The Earth collides with the red dot.}
	\label{fig:NEOForCE_scheme}
\end{figure}

\subsection{Combination of similar impactors}
\label{sec:combination_VIs}

On the last stage we check if some of the found VIs are essentially the same and point to the same possible collision. For this we integrate the orbits of each VI until the impact, saving their positions every 30 days. If at each integration  step several VIs are close to each other (distance $< 10^{-3}$~au), we take only the one with the highest impact probability value. We publish a possible collision if the impact probability is higher than $10^{-10}$. The reason why this threshold is significantly higher than $5\times10^{-8}$ threshold for the maximal impact probability is that if the maximal impact probability is too small it means the uncertainty of the asteroid around this area is too large. %On the other hand, if the maximal impact probability is not that small but the impact probability itself is indicates that the uncertainty region passes the Earth not through the center and small changes in the nominal orbit solution can fluctuate this impact probability.
On the other hand, if the maximal impact probability is not too small but the actual impact probability of a VI is very low, this means that the uncertainty region only grazes the Earth rather than passing through its center. In such cases, even small changes in the nominal orbital solution can lead to large fluctuations in the estimated impact probability, which is why these cases are interesting.

\subsection{Short summary of the method}
Below we provide a concise overview of the NEOForCE impact-monitoring algorithm.
A schematic illustration of the work is shown in Fig.~\ref{fig:NEOForCE_scheme}.

\begin{enumerate}

\item \textit{Sampling}. We create many virtual asteroids (VAs) along the Line of Variations (LOV) at the mean observation epoch (Section~\ref{sec:initial_sampling}).

\item \textit{Initial screening.} 
We make a first quick test to see whether the asteroid can ever come close to the Earth. For this, we compute the minimum MOID for a small group of VAs (Section~\ref{sec:initial_check}).

\item \textit{Propagation and close-approach detection.} We integrate the orbit of each VA into the future. We record every moment when the Earth comes close to the osculating orbit of a VA (osculating ellipse), together with the time and distance (Section~\ref{sec:pbm_initially}).

\item \textit{Curvilinear-region analysis.} For every case where the distance is smaller than 0.05~au, we build the curvilinear uncertainty region of the VA and compute its distance to the Earth. We also estimate the maximum impact probability that this VA could have (Section~\ref{sec:pbm_initially}).

\item \textit{Selection of representative scenarios.} From the previous step, we select only the most important situations: VAs that are neighbors along LOV and bracket the Earth at approximately the same time, have small distances between their uncertainty regions and the Earth, and show a noticeable impact probability (Section~\ref{sec:selection_representatives}).

\item \textit{Search for virtual impactors.} For each selected case, we try to improve the VA’s state vector so that the VA hits the Earth. We do this step by step, coming closer to an impact solution. When the VA enters the Earth’s sphere of influence at the probable collision time, we use the target-plane analysis (Sections~\ref{sec:vi_search} and Section~\ref{sec:target_plane}).

\item \textit{Combine similar impactors.} If two virtual impactors stay very close to each other on their way toward the Earth, we treat them as the same solution and keep the one with the higher impact probability (Section~\ref{sec:combination_VIs}).

\end{enumerate}

\section{Target plane analysis}
\label{sec:target_plane}

\subsection{Classical approach}

%The target plane analysis starts from the integration of the asteroid's orbit until the time it encounters the Earth. Let $(x_0,y_0,z_0,\dot{x_0},\dot{y_0},\dot{z_0})$ be the state vector of an asteroid at the epoch of observations, $t_0$, derived through orbital fitting, and let $\mathbf{C}_0$ be its variance-covariance matrix. Integrating the orbit of an asteroid with variational equations \citep{1964Battin_book} until the time at which it enters the sphere of influence of the Earth, $t$, provides the state vector $(x,y,z,\dot{x},\dot{y},\dot{z})$ at time $t$ and the partial derivative matrix,

The target plane (TP) analysis begins with the integration of the asteroid's orbit until its encounter with the Earth.  
Let $(x_0,y_0,z_0,\dot{x}_0,\dot{y}_0,\dot{z}_0)$ be the state vector of the asteroid at the observational epoch $t_0$, derived from orbital fitting, and let $\mathbf{C}_0$ be its variance–covariance matrix.  
Integrating the orbit with variational equations \citep{1964Battin_book} up to the time $t$ when the asteroid enters the Earth's sphere of influence yields the state vector $(x,y,z,\dot{x},\dot{y},\dot{z})$ and the partial derivative matrix

\begin{equation}
\mathbf{\Phi}(t_0,t) = \left(
\begin{array}{ccc}
\frac{\partial x }{\partial x_0} & \cdots &  \frac{\partial x }{\partial \dot{z}_0}\\
\vdots & \ddots & \vdots \\
\frac{\partial \dot{z} }{\partial x_0} & \cdots & \frac{\partial \dot{z} }{\partial \dot{z}_0} \\
\end{array}
\right) .
\label{eq:matr_Phi}
\end{equation}

The variance-covariance matrix at time $t$ is then

\begin{equation}
\mathbf{C}_{xyz\dot{x}\dot{y}\dot{z}} = \mathbf{\Phi}(t_0,t) \ \mathbf{C}_0 \ \mathbf{\Phi^\mathrm{T}}(t_0,t), 
\end{equation}

\noindent where $\mathrm{T}$ denotes the matrix transpose operation. Matrix $\mathbf{C}_{xyz\dot{x}\dot{y}\dot{z}}$ provides an approximation of the uncertainty region of an asteroid at time $t$.

%The velocity of an asteroid relative to the Earth, $\bm{v}_{rel}$ is almost identical for all VAs in this area, and they therefore all move in almost the same direction with an almost similar speed. The target plane by definition is the plane perpendicular to the relative velocity of two objects (the asteroid and the Earth in our case). The definition of the coordinate system on the target plane is arbitrary. In our system, we construct it the following:

The asteroid's relative velocity with respect to the Earth, $\bm{v}_{rel}$, is nearly the same for all virtual asteroids (VAs) in this region, meaning they share almost the same direction and speed. The target plane is defined as the plane perpendicular to $\bm{v}_{rel}$. The orientation of the coordinate axes on the plane is arbitrary; in our implementation we define them as

\begin{equation}
    \bm{\eta} = \frac{\bm{v}_{rel}}{||\bm{v}_{rel}||}; \ \bm{\xi} = \frac{\bm{\eta}\times\bm{z}}{||\bm{\eta}\times\bm{z}||}; \ \bm{\zeta} = \bm{\xi}\times\bm{\eta}
\end{equation}

\noindent where $\bm{\xi},\bm{\eta},\bm{\zeta}$ are the axis of the new system, with $\bm{\xi}$ and $\bm{\zeta}$ define the coordinates on the target plane and $\bm{z}$ is the z-axis in a Cartesian coordinate system. The center of the system is the asteroid.

%To estimate the impact probability, we first map this 6D uncertainty region onto the target plane. The VAs that collide with the Earth must have the distance from the Earth's center smaller than the Earth's radius.(see Fig.~\ref{fig:Target_plane}). Because of the gravitational focusing, however, the trajectory of an asteroid in the sphere of influence resembles a hyperbola, and the distance between the Earth and the asteroid is therefore smaller than if the asteroid moved in a straight unperturbed direction. This distance $q$ can be found from 

\begin{figure}
\centering
\includegraphics[width=0.7\columnwidth]{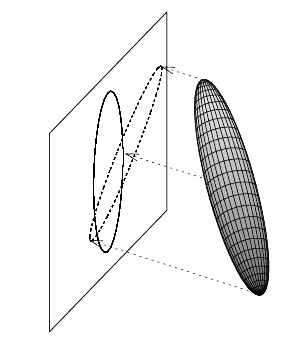}
  \caption{  {Schematic illustration of the target plane method. The ellipsoid represents the uncertainty region of an asteroid when it enters the sphere of influence, the plane is the target plane, the circle on the plane is the projection of the Earth, and the ellipse on the plane is the projection of the ellipsoid. Image credit: \citep{vavilov2018_iaaras}.}}
     \label{fig:Target_plane}
\end{figure}

To estimate impact probabilities, the 6D uncertainty region is projected onto the target plane. An asteroid will collide with the Earth if the distance from the Earth's center is smaller than its radius $R_\oplus$ (see Fig.~\ref{fig:Target_plane}).  
Because of gravitational focusing, however, the asteroid's trajectory inside the Earth's sphere of influence is hyperbolic, so the effective cross-section is larger. The minimum distance $q$ is given by

\begin{equation}
q = \frac{b}{\sqrt{1+\frac{v_s^2}{v_{\infty}^2}}}, 
\end{equation}

%\noindent where $v_s$ is the escape velocity from the surface of the Earth ($\approx$~11.186 $km \ s^{-1}$), $v_{rel}$ is the unperturbed geocentric velocity, and $b$ is the impact parameter, that is, distance on the target plane between the center of the Earth and the projection of the asteroid.

\noindent where $v_s \approx 11.186 \ \mathrm{km \ s^{-1}}$ is the Earth's surface escape velocity, {$v_{\inf}$ is the unperturbed geocentric velocity} and $b$ is the impact parameter (the distance on the target plane between the projections of the Earth’s and asteroid's centers).

%Any VA whose distance $q$ is smaller than the radius of the Earth, $R_{\oplus}$, will collide with the Earth, or equivalently, it will collide when its impact parameter $b$ is smaller than $R_{\oplus}\sqrt{1+\frac{v_s^2}{v_{\infty}^2}}$. Therefore, to estimate the collisional probability, we can map the uncertainty region onto the target plane and then compute the probability that an asteroid is closer to the projected center of the Earth than $R'_{\oplus} = R_{\oplus}\sqrt{1+\frac{v_s^2}{v_{\infty}^2}}$.

Thus, a collision occurs when $b < R_\oplus\sqrt{1+v_s^2/v_\infty^2}$. The effective collision radius is therefore $R'_\oplus = R_\oplus\sqrt{1+v_s^2/v_\infty^2}$.

%The transition from the 6D Cartesian coordinates and velocities to 2D coordinates on the target plane can be described as a multiplication by a $2\times6$ matrix, $\mathbf{B}$: $\bm{u} = \mathbf{B}\bm{w}$. The variance-covariance matrix on the target plane reads $\mathbf{C_{TP}} = \mathbf{B} \mathbf{C}_{xyz\dot{x}\dot{y}\dot{z}} \mathbf{B}^\mathrm{T}$ \citep{1952Raobook}. We placed the nominal asteroid in the center of the coordinate system and computed the impact probability as

The mapping from Cartesian state space (relative to the asteroid) to target-plane coordinates can be written as $\bm{u} = \mathbf{B}\bm{w}$ with a $2\times6$ matrix $\mathbf{B}$ and $\bm{w}$ being the relative to the nominal asteroid state vector. The covariance matrix on the target plane is then  
$\mathbf{C}_{TP} = \mathbf{B}\mathbf{C}_{xyz\dot{x}\dot{y}\dot{z}}\mathbf{B}^\mathrm{T}$ \citep{1952Raobook}.  
Placing the nominal solution at the origin, the impact probability reads

\begin{equation}
P = \frac{1}{2\pi |\mathrm{det}\mathbf{C_{TP}}|^{\frac{1}{2}}} \int \limits_{S_{R'_{\oplus}}} e^{-(\bm{u}^\mathrm{T} \mathbf{C_{TP}}^{-1} \bm{u})/2} \mathrm{d} \bm{u},
\end{equation}

\noindent where $S_{R'_{\oplus}}$ is a circle with a radius $R'_{\oplus}$.

\subsection{Beyond classical approach}

%The classical target plane approach is working very well until the relative velocity of the objects is small. Once the velocity is small the encounter does not happen fast, it takes several days and the motion of the Earth cannot be approximated as straightforward and the analysis can fail. This is the case for asteroid 2000~SG344 which motivated us to improve this approach. For a possible collision on September 18, 2069 our system found a VI, but the classical target plane method estimated the impact probability as zero, because the $(\xi, \zeta)$ coordinates of this VA on target plane, built when this VA entered the sphere of influence, was too far from Earth, even though this VA collided with the Earth several days later.

The classical target plane approach works well at typical encounter velocities. However, when the relative velocity is low, the encounter spans several days and the Earth's motion can no longer be approximated as rectilinear. In such cases, the method can be inaccurate. This problem arises for asteroid 2000~SG344: for a possible impact on 18 September 2069, NEOForCE identified a virtual impactor (VI), but the classical target plane method estimated zero probability because the projected $(\xi,\zeta)$ coordinates of the Earth (reminder that in our system the asteroid is in the center) at entry into the Earth's sphere of influence were too far from the center, despite an actual collision days later.

%We decided to implement a similar approach as described in \citep{2005Icar..173..362M}. First, we integrate the orbit of the asteroid further and indicate the timing and positions of the asteroid and the Earth at the closest distance. We build a modified target plane (MTP), which is a target plane perpendicular to the relative velocity at the closest distance and compute the coordinates on MTP: $\xi_{MTP},\zeta_{MTP}$. However, if the asteroid collides with the Earth we build MTP at the time of asteroid entering the Earth. Knowing the relative position and velocity of an asteroid at the closest distance we can compute the unperturbed velocity, $v_{\infty}$, (on the infinity in the two-body problem Earth-asteroid) and coordinates on the unperturbed classical target plane (TP2) by:

To address this, we adopted a modified approach similar to \citet{2005Icar..173..362M}. We extend the integration until the epoch of closest approach and construct a ''modified target plane (MTP)'' perpendicular to the relative velocity at that instant. If the asteroid collides with the Earth, the MTP is defined at the impact point. Using the relative position and velocity at closest approach, we compute the unperturbed velocity $v_\infty$ and derive the corresponding unperturbed target-plane coordinates $(\xi_{TP^*},\zeta_{TP^*})$ from the MTP coordinates $(\xi_{MTP},\zeta_{MTP})$:

\begin{equation}
    \begin{array}{l}
    E = \frac{v_{rel}^2}{2} - \frac{\mu}{r}; \ 
    v_{\infty} = \sqrt{2E}; \ 
    b = \frac{||\bm{r}\times \bm{v_{rel}}||}{v_{\infty}}; \  s = b / r \\
    \xi_{TP^*} = s \ \xi_{MTP}; \ \zeta_{TP^*} = s \ \zeta_{MTP}
    \end{array}
    \label{eq:TP1_to_TP2}
\end{equation}

%\noindent where $E$ is the total energy, $\mu=\frac{GM_{\oplus}}{R_{\oplus}}$, $M_{\oplus}$ and $R_{\oplus}$ are the mass and radius of the Earth. $b$ is the so-called impact parameter, and $\xi_{TP^*}$ and $\zeta_{TP^*}$ are coordinates on the classical target plane computed from modified target plane. It should be noted that a regular asteroid moves fast relative to the Earth at the encounter and the coordinates  $(\xi_{TP^*},\zeta_{TP^*})$ almost coincide with the coordinates $(\xi_{TP},\zeta_{TP})$ on the target plane constructed from the incoming to the sphere of influence (TP1). However, if the relative velocity is small the difference can be significant. 

\noindent where $E$ is the total orbital energy, $\mu = GM_\oplus/R_\oplus$, and $M_\oplus, R_\oplus$ are the Earth's mass and radius. For high-speed encounters, $(\xi_{TP^*},\zeta_{TP^*}) \approx (\xi_{TP},\zeta_{TP})$, but at low velocities the difference is significant.

%The only problem we have remaining is to map the uncertainty region on a target plane (and which one should we choose?). The only covariance matrix available for us is the one computed before entering the sphere of influence. We decided to use the following trick. We assume the transformation from TP1 to TP2 as rotation around the Earth on some angle and stretching by a scaling factor, $sc$. Both of them are estimated from :

Since the covariance matrix gives adequate estimation of the uncertainty region only before the entry into the sphere of influence, we approximate the mapping between TP1 (classical TP at entry) and TP2 (unperturbed TP built from MTP) as a rotation by an angle $\alpha$ and a scale factor $sc$, estimated from

\begin{equation}
    \begin{array}{l}
         sc = \sqrt{ \frac{\xi_{TP^*}^2 + \zeta_{TP^*}^2}{\xi_{TP}^2 + \zeta_{TP}^2} }  \\
         \alpha = \mathrm{atan2}(\zeta_{TP^*},\xi_{TP^*}) -   \mathrm{atan2}(\zeta_{TP},\xi_{TP}) .
    \end{array}
\end{equation}

For 2000~SG344, $sc$ can exceed 1.3 while $\alpha$ is only a couple of degrees. The relative velocity $\bm{v}_{\infty}$ is also significantly different from $\bm{v}_{rel}$, and hence the corrected gravitational focusing radius of the Earth, $R_{\oplus}'$ changes. That is why we sometimes had a situation where we found a VI but the $(\xi_{TP},\zeta_{TP})$ coordinates are further away than the corrected radius of the Earth, which of course led to wrong impact probability values.

%After mapping the covariance matrix $\mathbf{C}_{xyz\dot{x}\dot{y}\dot{z}}$ at the time of entering the sphere of influence on TP1, we scale it by factor $sc$ and turn it to angle $\alpha$ yielding 2$\times$2 matrix $\mathbf{C}_{TP}$. If the asteroid collides with the Earth the rotation by 2 degrees does not make any difference, however, if the asteroid is passing close but not colliding and we need to find a VI this rotation is extremely important to take into account.

We therefore map $\mathbf{C}_{xyz\dot{x}\dot{y}\dot{z}}$ onto TP1, rotate by $\alpha$, and scale by $sc$ to obtain the  $2\times2$ matrix $\mathbf{C}_{TP^*}$.

\subsection{Search for virtual impactor}

%To search for a VI and check where this possible collision is spurious we apply the method from \citep{2005Icar..173..362M}. The vector of Cartesian coordinates and velocities of the asteroid when asteroid enters the sphere of influence, $t$, can be split into two components $[\bm{r},\bm{s}]$, where $\bm{r}$  coincides with the coordinates on the target plane TP1 $(\xi_{TP},\zeta_{TP})$ and $\bm{s}$ is a 4D vector, which basically contains the distance to the target plane and the three components of the velocity $\bm{v}_{rel}$.

To confirm whether a potential encounter corresponds to a genuine collision, we follow the method of \citet{2005Icar..173..362M}. The 6D state vector at $t$ relative to the nominal state vector of the asteroid can be partitioned into $[\bm{r},\bm{s}]$, where $\bm{r}=(\xi_{TP},\zeta_{TP})$ are target-plane coordinates and $\bm{s}$ is the complementary 4D vector.

%Our first goal is to determine the 6D state vector $[\bm{\hat{r}},\bm{\hat{s}}]$ at a time $t$ of the orbit that leads to a collision and that has the highest value of the probability density function.
%Only vector $\bm{r}$ determines the collision with the Earth, and vector $\bm{s}$ is mapped onto the target plane. 
%First we compute $\bm{r}_*=(\xi_{TP^*},\zeta_{TP^*})$ by equation~\ref{eq:TP1_to_TP2}. Second, we compute covariance matrix on TP2, $\mathbf{C}_{TP^*}$, by rotating $\mathbf{C}_{TP}$ to angle $\alpha$ and scaling by $sc$. Third, we find $\bm{\hat{r}_*}$ on TP2 so that it is inside the Earth with radius $R_{\oplus}'$ and closest to $\bm{r}_*$ in terms of standard deviations, i.e. $\bm{\hat{r}}_*^{\mathrm{T}}  \mathbf{C}_{TP^*}^{-1} \bm{\hat{r}}_*$ is minimal. For instance, if the largest axis of the uncertainty region on TP2 intersects with the projection of the Earth we select the point on the largest axis. Then we transform it back to TP1 and find $\bm{\hat{r}}$.

We seek the adjusted state $[\hat{\bm{r}},\hat{\bm{s}}]$ corresponding to a collisional orbit with the highest probability density. The procedure is:  
1. Compute $\bm{r}_*=(\xi_{TP^*},\zeta_{TP^*})$ from Eq.~(\ref{eq:TP1_to_TP2}).  
2. Obtain $\mathbf{C}_{TP^*}$ by rotating and scaling $\mathbf{C}_{TP}$.  
3. Select $\hat{\bm{r}}_*$ inside $R'_\oplus$ that minimizes $\hat{\bm{r}}_*^\mathrm{T}\mathbf{C}_{TP^*}^{-1}\hat{\bm{r}}_*$.  
4. Transform back to TP1 to get $\hat{\bm{r}}$.  

The covariance matrix $\mathbf{C}$ of $[\bm{r},\bm{s}]$ at time $t$ can be decomposed as

\begin{equation}
    \mathbf{C} = 
    \left(
    \begin{array}{cc}
      \mathbf{C_{r}}   & \mathbf{C_{rs}} \\
       \mathbf{C_{sr}}  & \mathbf{C_s}
    \end{array}
    \right) .
\end{equation}

According to \citep{2005Icar..173..362M} , vector $\bm{s_*}$ can be found from

\begin{equation}
    \bm{s_*} = \bm{s} + \mathbf{C_r}^{-1} \mathbf{C_{sr}} (\bm{r}-\bm{r_*})
    \label{eq:s_star}.
\end{equation}

We then search for the the  vector $\bm{x}_{0*}$ at epoch of observations, $t_0$, that leads to vector $[\bm{r_*},\bm{s_*}]$ at time $t$, and hence, to the collision.
The linear relation reads

\begin{equation}
    [(\bm{r_*},\bm{s_*})] - [(\bm{r},\bm{s})] = \mathbf{W} \cdot \mathbf{\Phi}(t_0,t) \ \ ( \bm{\tilde{x}}_{0*}-\bm{x}_{0*} )
    \label{eq:x_star_from_s_star},
\end{equation}

%\noindent where matrix   {$\mathbf{W}$ is a $6\times6$} rotation matrix leading to the coordinate system related to the target plane. As in Section~\ref{sec:vi_search} we also consider $\frac{3}{4}( \bm{\tilde{x}}_{0*}-\bm{x}_{0*} )$ and $\frac{1}{2}( \bm{\tilde{x}}_{0*}-\bm{x}_{0*} )$ and select the closest one to the Earth at time $t$. If all these tree are more distant than $\bm{x}_{0*}$ at time $t$ we take $\frac{1}{6}( \bm{\tilde{x}}_{0*}-\bm{x}_{0*} )$.

\noindent where $\mathbf{W}$ is the $6\times6$ rotation matrix into the TP2 frame. As in Section~\ref{sec:vi_search} we also consider $\frac{3}{4}( \bm{\tilde{x}}_{0*}-\bm{x}_{0*} )$ and $\frac{1}{2}( \bm{\tilde{x}}_{0*}-\bm{x}_{0*} )$ and select the closest one to the Earth at time $t$. If all these three are more distant than $\bm{x}_{0*}$ at time $t$ we take $\frac{1}{10}( \bm{\tilde{x}}_{0*}-\bm{x}_{0*} )$.

If the state vector $\bm{\tilde{x}}_{0*}$ does not lead to a collision approximately at time $t$, this process needs to be iterated {taking $\bm{\tilde{x}}_{0*}$ as $\bm{{x}}_{0*}$ and using Eqs.~(\ref{eq:s_star}) and (\ref{eq:x_star_from_s_star})}. If the procedure does not determine a collision after 10 iterations or it is outside of 9-$\sigma$ 6D ellipsoid, we consider this impactor as spurious.

\section{Practical realization and discussion}
\label{sec:results}

In this version of the system we used 
\Dima{collocational integrator Lobbie of 14-th order \citep{2022SoSyR..56...32A} for integrating the orbits.} %\citet{1974CeMec..10...35E} integrator of 15-th order. 
In the model we included gravitational influence of 8 major planets (and dwarf planet Pluto, for historical reasons), {as well as post-Newtonian formalism for the Sun only}. We also considered the Earth and the Moon separately. %The coordinates and velocities of the planets are obtained from planetary ephemeris DE440 \citep{2021AJ....161..105P}. 
The orbits of asteroids and their covariance matrices are taken from DynAstVO database\footnote{\url{https://epn.imcce.fr/dynastvo.html}} \citet{desmars17} \Dima{for the automated impact monitoring}.

\subsection{Completeness problem}
\label{sec:completeness_problem}

The completeness of our system is determined by the threshold of impact probability that can be detected certainly.
Following \citep{2005Icar..173..362M} one can estimate the \textit{generic completion level}: the highest impact probability that can be missed by our monitoring system if the uncertainty trail on the target plane is fully linear. 

Under these assumptions, a possible collision can be missed if both representatives found in Section~\ref{sec:selection_representatives} are further than 1~au from the Earth. This literally means that the distance between two subsequent VAs at the time of this possible collision exceeds 2~au along the orbit. 
%We take  $10^{-4}$~au as the diameter of the Earth: slightly higher than the real value ($8.54\cdot10^{-5}$~au) in order to take into account the gravitational focusing of the Earth. 
According to the sampling of VAs along the LOV, the maximal integral probability of a Gaussian distribution on the interval between two subsequent VAs is approximately $3\times10^{-4}$. This is slightly higher than if we sampled VAs along the LOV according to Gaussian distribution. In this case, the probability would be $2\times10^{-4}$. Our objective is hence the following: If we took a sample of 5001 VAs according to the Gaussian distribution, the latest VAs would be only 3.54\,sigmas away from the nominal state vector ($l_{2500} = 3.54$ from equation~\ref{eq:sampling}). By sampling the Laplace distribution $l_{2500} = 5.53$ instead, we have VAs much further from the nominal state vector, without loosing much in terms of completeness.
Hence, the highest impact probability that can be missed is approximately:

\begin{equation}
    \frac{D_{\oplus}[au]}{2~\mathrm{au}} 3\times10^{-4} %\in [1.5 \cdot 10^{-8} , 1.7\cdot 10^{-7}]
\end{equation}

To take into account the gravitational focusing of the Earth in order to properly estimate the completeness we need to increase the diameter of the Earth $D_{\oplus}$ by $\sqrt{1 + v_e^2/v_{\infty}^2}$ \citep{1993BAAS...25.1236C}, where $v_e$ is the escape velocity from the Earth's surface ($\approx 11.186$~km/s) and $v_{\infty}$ is the unperturbed geocentric velocity (approximately the velocity when crossing the sphere of influence). When the relative velocity of an asteroid and the Earth is high, the  diameter of the Earth doesn't change much and is about $10^{-4}$, which makes the completeness to be around $1.5\times10^{-8}$. However, for a very slow moving asteroids (with 1~km/s relative velocity) the diameter is 11.22 times larger, that lowers the completeness to $1.7\times 10^{-7}$. 
In Section~\ref{sec:selection_representatives} we exclude all the representatives with estimated maximal impact probability lower than $5\times 10^{-8}$, which sounds reasonable. %In this case, we are more or less sure that the two subsequent VAs chosen in Section~\ref{sec:selection_representatives} made the same number of revolutions around the Sun while going to the time of this possible impact. 

\subsection{Comparison to Sentry}

To evaluate the robustness of our system, we performed detailed comparisons using five representative asteroids: 2005~QK76, 2008~JL3, 2023~DO, 2008~EX5, and 2000~SG344. We selected these asteroids from the web-page Sentry of Central for Near-Earth Objects Study (CNEOS)\footnote{https://cneos.jpl.nasa.gov/sentry}. We were looking for objects at the top of the list with high cumulative impact probability and significant number of possible impacts with probabilities higher than $5\times10^{-7}$ in order to be able to make a comparison. The asteroids have different observational arcs (from 1.7 days to 1.4 years), and different number of observations (from 14 to 52).
For each case, we adopted the same orbital solutions {and the same covariance matrices} as those used by NASA’s Sentry-II system, retrieved via the NASA SSD API\footnote{\url{https://ssd-api.jpl.nasa.gov}}
 on September 9, 2025, so that the results would be directly comparable.

The outcomes of our system and Sentry-II are presented in the tables below. In each table, “Date” refers to the epoch of the possible collision identified by NEOForCE, while $P_{{NEOForCE}}$ and $P_{{NASA}}$ are the corresponding impact probabilities computed by our system and Sentry-II, respectively. We attempted to match virtual impactors (VIs) such that, for collisions occurring at similar epochs, both the probability and the distance metric $\sigma_{VI}$ are comparable.

$\sigma_{VI}$ quantifies how far a VI lies from the nominal orbit in terms of standard deviations and is defined as

\begin{equation}
\sigma_{VI}^2 = [\bm{x}_{0*} - \bm{x_0}]^{\mathrm{T}} \mathbf{C}_0^{-1} [\bm{x}_{0*} - \bm{x_0}],
\label{eq:sigma_vi}
\end{equation}

\noindent where $\bm{x}_{0*}$ is the state vector of the VI, $\bm{x}_0$ is the nominal state vector, and $\mathbf{C}_0$ is the covariance matrix. %Sentry-II estimates $\sigma{VI}$ using the root-mean-square residuals of the observational fit for the impacting solution.

It is important to note that similar $\sigma_{VI}$ values do not necessarily imply that the same VI was found: many virtual asteroids can lie on the same 5D ellipsoidal surface defined by a given $\sigma_{VI}$. Conversely, two very similar VIs may have different $\sigma_{VI}$ values, for example, a small shift along the direction of the smallest eigenvector of $\mathbf{C}_0$ can leave the overall trajectory nearly unchanged while significantly altering $\sigma_{VI}$.

\begin{table}[h]
\caption{Comparison table for asteroid 2008~EX5.}
\label{tab:2008_EX5}
\begin{tabular}{lllll}
Date & $P_{NEOForCE}$ & $\sigma_{VI}$ & $P_{NASA}$ & $\sigma_{VI_{NASA}}$ \\
\hline
2056-10-8.9 & 7.04e-08 & 0.504 & 7.00e-08 & 1.028 \\
\hline
2059-10-9.5 & 5.38e-08 & 0.505 &  &  \\
\hline
2061-10-9.0 & 3.00e-08 & 0.501 & 3.00e-08 & 1.400 \\
\hline
2062-10-9.5 & 6.64e-07 & 0.562 & 7.80e-07 & 0.604 \\
\hline
2063-10-9.5 & 5.09e-08 & 0.498 & 4.80e-08 & 0.618 \\
\hline
2065-10-9.0 & 4.41e-08 & 0.506 & 4.70e-09 & 0.564 \\
2065-10-09.09 &  &  & 4.50e-08 & 1.085 \\
\hline
2067-10-9.6 & 4.11e-08 & 0.503 &  &  \\
2067-10-9.8 & 9.43e-08 & 0.710 &  &  \\
\hline
2068-10-9.0 & 5.09e-07 & 0.551 & 6.90e-07 & 0.853 \\
2068-10-08.85 &  &  & 3.50e-10 & 2.175 \\
\hline
2069-10-9.1 & 1.02e-07 & 0.255 & 1.00e-07 & 1.324 \\
\hline
2070-10-09.39 &  &  & 6.20e-08 & 0.649 \\
\hline
2071-10-9.7 & 2.10e-06 & 0.371 & 2.00e-06 & 0.644 \\
2071-10-09.63 &  &  & 3.40e-08 & 1.238 \\
\hline
2072-10-9.0 & 1.94e-05 & 0.919 & 2.70e-05 & 0.922 \\
2072-10-8.9 & 1.37e-06 & 1.533 & 1.40e-06 & 1.685 \\
\hline
2074-10-9.5 & 4.62e-07 & 0.481 & 4.80e-07 & 0.592 \\
2074-10-9.4 & 1.26e-07 & 1.615 & 1.30e-07 & 1.707 \\
\hline
2075-10-9.8 & 2.19e-08 & 0.917 & 6.70e-10 & 1.449 \\
\hline
2076-10-8.9 & 6.60e-08 & 0.394 & 6.30e-08 & 0.447 \\
2076-10-8.9 & 2.52e-08 & 0.420 &  &  \\
\hline
2077-10-9.1 & 3.11e-07 & 0.542 & 3.50e-07 & 0.699 \\
\hline
2079-10-9.8 & 3.74e-08 & 0.266 & 3.80e-07 & 0.286 \\
2079-10-9.6 & 3.45e-08 & 0.578 & 3.40e-08 & 1.161 \\
\hline
2080-10-9.0 & 5.28e-08 & 0.715 &  &  \\
\hline
2081-10-9.1 & 8.56e-08 & 1.579 &  &  \\
\hline
2082-10-9.3 & 9.08e-08 & 1.620 & 9.30e-08 & 2.029 \\
2082-10-9.3 & 1.80e-10 & 0.576 &  &  \\
\hline
2083-10-9.7 & 1.89e-05 & 0.844 & 2.00e-05 & 1.002 \\
2083-10-09.66 &  &  & 1.30e-08 & 0.617 \\
\hline
2084-10-8.9 & 5.56e-08 & 1.582 & 5.40e-08 & 1.619 \\
\hline
2086-10-9.5 & 4.42e-10 & 0.257 &  &  \\
\hline
2088-10-09.02 &  &  & 5.40e-09 & 1.445 \\
\hline
2090-10-9.5 & 5.44e-08 & 0.030 & 5.80e-08 & 1.282 \\
\hline
2093-10-09.25 &  &  & 2.30e-09 & 0.045 \\
\hline
\end{tabular}
\end{table}

%\subsubsection{Asteroid 2025~JU}

%Asteroid 2025~JU is newly discovered, with 38 observations spanned over 35 days. Its cumulative impact probability is about $6\times10^{-5}$, not very high and it also has only a few significant possible impacts. Table~\ref{tab:2025_JU} shows an iconic agreement between two systems with Sentry loosing three possible collisions with a probability less than $5.3\times10^{-9}$ (way below the mentioned completeness level).

\subsubsection{Asteroid 2008~EX5}

{Asteroid 2008~EX5 is newly discovered, with 61  observations spanned over 32 days. Its cumulative impact probability is about $4.5\times10^{-5}$, not very high and it also has only a few significant possible impacts. Table~\ref{tab:2008_EX5} shows an  agreement between two systems with Sentry loosing 8 possible collisions with a probability less than $1\times10^{-7}$ (below the mentioned completeness level) and NEOForCE looses 7 possible collisions with probabilities less than $5\times10^{-8}$, below the threshold for maximal impact probability.}

\begin{table}[h]
\caption{Comparison table for asteroid 2005~QK76.}
\label{tab:2005_QK76}
\begin{tabular}{lllll}

Date & $P_{NEOForCE}$ & $\sigma_{VI}$ & $P_{NASA}$ & $\sigma_{VI_{NASA}}$ \\
\hline
2030-02-26.3 & 4.59e-05 & 0.156 & 3.80e-05 & 0.263 \\
\hline
2038-02-26.3 & 1.63e-05 & 1.179 & 1.60e-05 & 2.314 \\
2038-02-26.3 & 1.20e-08 & 2.556 & 7.10e-09 & 5.534 \\
2038-02-26.2 & 1.71e-07 & 1.742 &  &  \\
\hline
2041-02-26.0 & 6.40e-08 & 0.707 &  &  \\
2041-02-26.0 & 1.66e-08 & 0.706 &  &  \\
\hline
2048-02-26.8 & 3.41e-06 & 0.662 & 3.40e-06 & 3.625 \\
2048-02-26.8 & 1.33e-05 & 0.604 & 1.30e-05 & 3.786 \\
\hline
2059-02-26.5 & 5.58e-10 & 4.127 & 1.60e-10 & 6.973 \\
\hline
2062-02-26.2 & 3.97e-08 & 2.220 &  &  \\
\hline
2078-02-26.2 & 2.24e-10 & 2.924 &  &  \\
\hline
2083-02-26.5 & 4.49e-08 & 3.123 &  &  \\
\hline
\end{tabular}

\end{table}

\subsubsection{Asteroid 2005~QK76}

Asteroid 2005~QK76 has been discovered on August 30 2025. It has only 14 observations spanned over 1.67 days. Such a short observation arc makes its orbit highly uncertain and a good test for comparison. %Sentry-II does not use the covariance matrix only the observations itself, while NEOForCE uses the covariance matrix in coordinates and velocities at the mean epoch of observations. 
As one can see from Table~\ref{tab:2005_QK76}, there is no significant difference in the impact probabilities of the possible collisions found. However, the values of $\sigma_{VI}$ are substantially different, with larger values for Sentry-II. 
Sentry-II did not find the possible collision on February 26, 2038 with impact probability of $1.7\times10^{-7}$, but this is at the completeness limit of Sentry-II.

\begin{table}
\caption{Comparison table for asteroid 2023~DO.}
\label{tab:2023_DO}
\begin{tabular}{lllll}

Date & $P_{NEOForCE}$ & $\sigma_{VI}$ & $P_{NASA}$ & $\sigma_{VI_{NASA}}$ \\
\hline
2057-03-23.8 & 4.38e-04 & 0.690 & 4.40e-04 & 0.693 \\
\hline
2058-03-24.2 & 4.47e-07 & 2.549 & 4.80e-07 & 2.555 \\
2058-03-23.8 & 1.51e-08 & 0.689 &  &  \\
\hline
2059-03-24.0 & 8.90e-06 & 0.714 & 8.40e-06 & 0.717 \\
\hline
2060-03-23.3 & 5.12e-07 & 0.692 &  &  \\
\hline
2061-03-24.0 & 8.42e-08 & 2.244 &  &  \\
2061-03-23.9 & 3.92e-09 & 2.186 &  &  \\
2061-03-23.8 & 6.65e-09 & 2.137 &  &  \\
\hline
2062-03-24.0 & 8.05e-07 & 2.797 & 6.90e-07 & 2.803 \\
2062-03-23.8 & 2.05e-07 & 0.683 & 2.10e-07 & 0.687 \\
\hline
2063-03-24.4 & 6.52e-08 & 2.242 &  &  \\
\hline
2064-03-23.3 & 8.18e-08 & 0.696 & 7.90e-08 & 0.706 \\
2064-03-23.4 & 6.05e-07 & 0.679 & 5.90e-07 & 0.680 \\
2064-03-23.6 & 3.50e-08 & 1.413 &  &  \\
\hline
2065-03-23.9 & 2.59e-08 & 2.181 & 7.30e-08 & 2.245 \\
2065-03-23.9 & 4.75e-06 & 1.780 & 4.80e-06 & 1.785 \\
\hline
2066-03-24.1 & 2.12e-08 & 1.412 & 2.00e-08 & 1.417 \\
2066-03-23.8 & 2.05e-07 & 0.697 & 1.90e-07 & 0.702 \\
2066-03-23.9 & 1.25e-06 & 0.673 & 1.10e-06 & 0.674 \\
2066-03-24.05 &  &  & 1.50e-09 & 2.388 \\
\hline
2067-03-24.3 & 2.61e-08 & 2.192 & 2.00e-08 & 2.553 \\
2067-03-24.4 & 1.40e-07 & 2.149 & 1.30e-07 & 2.154 \\
2067-03-24.4 & 8.86e-08 & 2.292 &  &  \\
2067-03-24.4 & 6.91e-08 & 2.180 &  &  \\
\hline
2068-03-23.4 & 7.52e-07 & 0.664 & 8.60e-07 & 0.670 \\
2068-03-23.1 & 9.42e-09 & 0.713 &  &  \\
\hline
2069-03-23.8 & 7.34e-08 & 2.128 & 7.00e-08 & 2.133 \\
2069-03-23.5 & 2.13e-09 & 0.679 &  &  \\
\hline
2070-03-23.6 & 1.30e-08 & 0.713 &  &  \\
\hline
2071-03-24.2 & 3.77e-09 & 2.795 &  &  \\
\hline
2072-03-23.6 & 6.97e-08 & 2.410 & 6.80e-08 & 2.416 \\
2072-03-23.5 & 3.71e-08 & 2.452 &  &  \\
2072-03-23.5 & 4.04e-10 & 2.257 &  &  \\
2072-03-23.5 & 2.43e-10 & 2.190 &  &  \\
2072-03-23.5 & 1.29e-09 & 1.422 &  &  \\
2072-03-23.5 & 8.30e-09 & 1.413 &  &  \\
2072-03-22.9 & 7.58e-09 & 0.713 &  &  \\
\hline
2073-03-23.8 & 4.47e-07 & 2.666 & 2.90e-07 & 2.671 \\
\hline
2074-03-24.1 & 9.31e-09 & 2.408 &  &  \\
2074-03-24.1 & 1.03e-09 & 2.383 &  &  \\
2074-03-24.0 & 9.12e-09 & 1.838 &  &  \\
\hline
2076-03-23.6 & 1.62e-10 & 2.458 &  &  \\
2076-03-23.7 & 8.15e-09 & 2.248 &  &  \\
2076-03-23.7 & 3.38e-08 & 1.418 &  &  \\
\hline
2077-03-23.4 & 3.40e-08 & 0.664 & 3.40e-08 & 0.667 \\
2077-03-23.3 & 1.69e-09 & 0.714 &  &  \\
\hline
2078-03-23.9 & 4.39e-09 & 1.422 & 7.90e-09 & 2.306 \\
\hline
2080-03-23.44 &  &  & 3.30e-08 & 1.841 \\
2080-03-23.57 &  &  & 1.20e-08 & 2.154 \\
\hline
2082-03-24.20 &  &  & 6.70e-08 & 2.252 \\
\hline
2085-03-23.6 & 1.79e-06 & 2.221 &  &  \\
\hline
2092-03-23.90 &  &  & 2.70e-09 & 1.385 \\
\hline
\end{tabular}

\end{table}

\subsubsection{Asteroid 2023 DO}

%This asteroid has a relative longer observation span of 41 days and 52 observations. It was discovered on February 13, 2023 and has 25 published possible collision by Sentry-II. The comparison table~\ref{tab:2023_DO} shows that our NEOForCE system found 42 possible collisions in total. NEOForCE did not find 6 possible collisions mentioned by Sentry-II with impact probabilities below $6.7\cdot10^{-8}$, which is close to our completeness threshold of $5\cdot10^{-8}$. Sentry-II did not find  23 possible collisions mentioned by our system, with maximal impact probability of $2.2\cdot10^{-6}$ on March 23, 2085, one collision with probability $6.3\cdot10^{-7}$ on March 23, 2060, and $9.1\cdot10^{-8}$ on March 24, 2061.

This asteroid was observed over a relatively longer arc of 41 days, with 52 recorded observations. Discovered on February 13, 2023, it has 25 possible collisions published by Sentry-II. As shown in Table~\ref{tab:2023_DO}, our NEOForCE system identified 47 possible collisions in total. Of these, NEOForCE did not reproduce 5 solutions listed by Sentry-II, all with impact probabilities below $6.7\times10^{-8}$, close to our completeness threshold of $5\times10^{-8}$. Conversely, Sentry-II missed 27 possible collisions detected by NEOForCE, including one with a probability of $1.79\times10^{-6}$ on March 23, 2085, another at $5.1\times10^{-7}$ on March 23, 2060, and one at $8.4\times10^{-8}$ on March 24, 2061.

\subsubsection{Asteroid 2008~JL3}

%This asteroid has a small observation arc as well, 7 days, and only 35 observations. It was discovered on May 2, 2008 and but has 44 possible collisions found by Sentry-II. The comparison of our system with Sentry-II on Table~\ref{tab:2008_JL3} reveals, that NEOForCE did not find 5 possible collisions with impact probabilities below $1.6\cdot10^{-8}$, which is below our estimated completeness limit. Sentry-II on the other hand did not find 42 possible collisions with highest impact probability of $1.11\cdot10^{-6}$ on April 28, 2109. It also missed 4 possible collisions with an impact probability higher $10^{-7}$.

This asteroid was observed over a short arc of only 7 days, yielding 35 measurements in total. Discovered on May 2, 2008, it has 44 possible collision solutions listed in Sentry-II. A comparison of our system with Sentry-II (Table~\ref{tab:2008_JL3}) shows that NEOForCE identified 87 possible collisions but did not recover 3 mentioned by Sentry-II, all with impact probabilities below $4.6\times10^{-9}$ --- way below our completeness threshold. In contrast, Sentry-II missed 46 possible collisions, including one with the highest reported impact probability of $1.26\times10^{-6}$ on April 28, 2109, as well as three others with probabilities exceeding $10^{-7}$.

\subsubsection{Asteroid 2000~SG344}

\begin{table}
\caption{Comparison table for asteroid 2000~SG344 (beginning).}
\label{tab:2000_SG344_beginning}
\begin{tabular}{lllll}

Date & $P_{NEOForCE}$ & $\sigma_{VI}$ & $P_{NASA}$ & $\sigma_{VI_{NASA}}$ \\
\hline
2069-09-18.5 & 1.96e-07 & 4.008 & 2.20e-07 & 3.995 \\
\hline
2070-09-17.3 & 2.18e-04 & 1.732 & 2.30e-04 & 1.727 \\
\hline
2071-09-16.0 & 1.03e-03 & 0.864 & 1.00e-03 & 0.862 \\
2071-09-10.4 & 1.68e-04 & 0.720 & 1.40e-04 & 0.717 \\
\hline
2072-02-3.5 & 3.59e-06 & 1.838 & 2.70e-06 & 1.832 \\
2072-09-13.8 & 1.59e-06 & 3.860 & 1.40e-06 & 3.845 \\
2072-09-8.5 & 5.06e-07 & 3.683 &  &  \\
\hline
2073-02-1.7 & 1.44e-05 & 0.737 & 1.30e-05 & 0.735 \\
2073-02-12.0 & 1.04e-05 & 1.790 & 1.00e-05 & 1.782 \\
2073-02-9.2 & 2.06e-06 & 1.834 & 5.00e-06 & 1.827 \\
2073-01-29.26 &  &  & 6.70e-07 & 1.839 \\
\hline
2074-02-10.5 & 4.02e-05 & 0.787 & 4.10e-05 & 0.785 \\
2074-02-8.5 & 4.48e-05 & 0.748 & 4.00e-05 & 0.745 \\
2074-01-28.5 & 4.43e-06 & 0.739 & 3.40e-06 & 0.736 \\
\hline
2078-09-1.6 & 2.19e-05 & 0.880 & 2.30e-06 & 0.876 \\
2078-09-7.0 & 5.66e-07 & 1.719 &  &  \\
\hline
2079-08-13.9 & 1.44e-06 & 0.888 & 2.10e-06 & 0.886 \\
2079-09-9.8 & 2.36e-07 & 0.871 & 1.20e-08 & 0.719 \\
2079-09-8.5 & 5.64e-07 & 0.862 & 1.30e-06 & 0.881 \\
2079-09-8.5 & 4.58e-07 & 1.716 & 2.60e-06 & 1.647 \\
2079-07-30.2 & 2.95e-07 & 1.719 &  &  \\
2079-09-3.4 & 6.61e-06 & 0.888 &  &  \\
\hline
2080-09-6.2 & 1.62e-06 & 0.891 & 1.30e-06 & 0.888 \\
2080-09-8.5 & 4.62e-07 & 1.712 & 4.20e-07 & 1.706 \\
2080-09-7.6 & 3.28e-07 & 0.870 &  &  \\
2080-09-6.4 & 2.51e-07 & 0.862 &  &  \\
2080-09-2.8 & 4.34e-07 & 0.857 &  &  \\
2080-09-6.7 & 3.32e-08 & 1.735 &  &  \\
\hline
2081-09-7.6 & 2.62e-06 & 0.895 & 1.60e-06 & 0.892 \\
2081-09-10.1 & 5.07e-07 & 1.709 & 4.00e-07 & 1.703 \\
2081-09-7.0 & 1.37e-07 & 0.870 &  &  \\
2081-09-3.0 & 4.14e-07 & 0.852 &  &  \\
2081-08-30.9 & 2.74e-08 & 0.720 &  &  \\
2081-09-4.1 & 4.73e-08 & 1.745 &  &  \\
\hline
2082-09-9.6 & 5.60e-07 & 1.735 & 1.80e-06 & 1.321 \\
2082-07-5.0 & 3.17e-07 & 0.713 &  &  \\
2082-06-29.1 & 4.71e-08 & 1.857 &  &  \\
2082-09-7.5 & 2.41e-06 & 0.898 &  &  \\
2082-09-7.3 & 1.45e-07 & 0.869 &  &  \\
2082-09-3.3 & 4.58e-07 & 0.848 &  &  \\
2082-09-4.5 & 1.06e-07 & 1.745 &  &  \\
\hline
2083-09-7.2 & 2.94e-06 & 0.902 & 1.90e-06 & 0.904 \\
2083-09-3.6 & 4.34e-07 & 0.847 & 3.90e-07 & 1.218 \\
2083-09-9.5 & 6.60e-07 & 1.703 & 5.30e-07 & 2.433 \\
2083-01-31.8 & 2.22e-09 & 0.747 &  &  \\
2083-01-30.5 & 5.14e-09 & 1.844 &  &  \\
2083-06-23.4 & 5.43e-08 & 1.857 &  &  \\
2083-09-7.2 & 8.31e-08 & 0.869 &  &  \\
2083-08-31.6 & 1.96e-07 & 0.712 &  &  \\
2083-09-5.2 & 9.63e-08 & 1.756 &  &  \\
2083-08-32.3 & 5.39e-09 & 1.853 &  &  \\
\end{tabular}

\end{table}

Asteroid 2000~SG344 turns out to be a hard test for our systems. It was discovered on 29 September 2000 by the LINEAR survey at Socorro, New Mexico. The orbital solution  used today by NASA JPL is based on observations spanning from 15 May 1999 to 3 October 2000, giving it an observational arc of about 1.4 years, with only a modest number of measurements available for orbit fitting (31 observations). Its orbital period is nearly one year, the inclination relative to the ecliptic is almost zero, and the aphelion lies just beyond Earth’s orbit. These parameters result in very low encounter velocities with Earth (about 1.4~km/s) and a continuum of predicted close approaches after {several resonance returns} after 2070. Because of this unusual dynamical configuration, 2000~SG344 has been the subject of a  dedicated impact-hazard study \citep{2001DDA....32.0803C}.

Sentry-II system indicates 300 possible impacts of this asteroids with the Earth, from which 138 with probabilities higher than $10^{-6}$. NEOForCE system found 1363 possible impacts and 172 with probabilities $>10^{-6}$. 

{
For the reasons discussed above, the computation time for asteroid 2000~SG344 was significantly higher (more then 6x longer than for 2008~JL3). 
Table~\ref{tab:computation_times} reports the total computation time, obtained in parallel mode using two threads on an Intel(R) Xeon(R) Gold 6230 CPU at 2.10\,GHz, together with the number of situations in which the virtual impactor (VI) search procedure was triggered.
}

\medskip

\begin{center}
%\vspace{3ex}
\tablecaption{Comparison table for asteroid 2008 JL3}

\tablefirsthead{Date & $P_{NEOForCE}$ & $\sigma_{VI}$ & $P_{NASA}$ & $\sigma_{VI_{NASA}}$ \\ \hline}
\tablehead{Date & $P_{NEOForCE}$ & $\sigma_{VI}$ & $P_{NASA}$ & $\sigma_{VI_{NASA}}$ \\ \hline}
%\tabletail{}
\label{tab:2008_JL3}

\begin{supertabular}{lllll}
2027-04-31.3 & 1.39e-04 & 0.111 & 1.50e-04 & 0.120 \\
\hline
2030-04-30.8 & 3.00e-08 & 4.242 & 3.00e-08 & 4.286 \\
\hline
2034-04-30.5 & 1.67e-07 & 0.105 & 1.60e-07 & 0.110 \\
\hline
2043-05-01.05 &  &  & 4.60e-09 & 0.117 \\
\hline
2047-05-01.52 &  &  & 3.20e-09 & 0.116 \\
\hline
2048-04-30.5 & 1.32e-08 & 0.115 &  &  \\
\hline
2051-04-31.4 & 7.34e-08 & 0.104 &  &  \\
\hline
2061-04-31.1 & 5.74e-09 & 0.115 &  &  \\
\hline
2063-05-2.0 & 2.28e-07 & 0.092 & 2.10e-07 & 0.096 \\
\hline
2065-04-27.7 & 8.02e-08 & 0.911 & 7.90e-08 & 0.920 \\
\hline
2071-04-30.3 & 2.63e-06 & 1.022 & 2.60e-06 & 1.032 \\
\hline
2072-04-30.5 & 2.03e-08 & 0.117 &  &  \\
\hline
2075-05-2.1 & 8.78e-08 & 1.598 & 1.00e-07 & 1.614 \\
2075-05-2.0 & 9.28e-08 & 0.081 &  &  \\
\hline
2077-04-28.5 & 9.27e-09 & 2.222 &  &  \\
\hline
2080-04-26.93 &  &  & 2.40e-09 & 0.193 \\
\hline
2081-04-31.5 & 1.39e-08 & 0.013 &  &  \\
\hline
2082-04-27.4 & 4.08e-09 & 0.084 & 4.60e-09 & 0.153 \\
\hline
2083-05-1.8 & 1.60e-08 & 0.095 &  &  \\
\hline
2087-04-31.2 & 2.03e-08 & 2.145 &  &  \\
\hline
2090-05-2.2 & 7.96e-08 & 0.099 & 7.70e-08 & 0.098 \\
2090-05-1.6 & 1.25e-09 & 0.158 & 5.80e-09 & 0.167 \\
2090-05-1.6 & 3.11e-07 & 0.193 & 2.80e-07 & 0.195 \\
2090-04-31.5 & 5.40e-07 & 0.195 & 5.50e-07 & 0.197 \\
2090-04-31.2 & 6.47e-07 & 0.229 & 5.80e-07 & 0.230 \\
2090-04-31.5 & 5.52e-07 & 1.958 & 5.30e-07 & 1.978 \\
2090-04-31.1 & 2.27e-08 & 0.188 &  &  \\
2090-05-1.5 & 1.14e-08 & 2.291 &  &  \\
2090-05-1.8 & 7.40e-08 & 2.792 &  &  \\
2090-05-1.9 & 4.76e-08 & 3.011 &  &  \\
\hline
2091-04-28.8 & 9.86e-09 & 0.919 & 9.10e-09 & 0.923 \\
\hline
2092-04-28.9 & 2.72e-08 & 1.041 & 3.10e-08 & 1.052 \\
2092-04-28.5 & 2.75e-08 & 0.189 &  &  \\
\hline
2093-05-1.6 & 1.32e-08 & 0.166 &  &  \\
\hline
2095-04-30.8 & 1.71e-08 & 0.085 & 1.60e-08 & 0.092 \\
2095-04-31.4 & 3.21e-08 & 0.139 & 2.70e-08 & 0.139 \\
\hline
2098-04-31.0 & 1.45e-08 & 0.190 &  &  \\
\hline
2099-04-31.0 & 6.33e-09 & 2.141 &  &  \\
2099-04-30.8 & 1.53e-08 & 0.084 &  &  \\
2099-04-30.9 & 1.53e-08 & 1.023 &  &  \\
\hline
2100-04-27.8 & 1.87e-07 & 1.267 & 1.80e-07 & 1.283 \\
2100-04-27.9 & 2.63e-06 & 0.264 & 2.50e-06 & 0.275 \\
2100-04-27.9 & 8.81e-07 & 0.174 & 8.60e-07 & 0.176 \\
2100-04-27.8 & 1.25e-07 & 0.131 & 1.10e-07 & 0.134 \\
2100-04-27.4 & 9.47e-10 & 2.651 &  &  \\
2100-04-27.8 & 1.08e-08 & 2.297 &  &  \\
\hline
2102-05-1.8 & 1.51e-08 & 0.231 & 3.40e-09 & 2.267 \\
\hline
2103-04-28.3 & 4.90e-08 & 2.214 &  &  \\
2103-04-28.5 & 1.86e-09 & 0.096 &  &  \\
\hline
2109-04-27.9 & 2.55e-07 & 0.594 & 2.30e-07 & 0.601 \\
2109-04-28.1 & 6.88e-08 & 1.089 & 6.50e-08 & 1.103 \\
2109-04-28.1 & 4.91e-09 & 0.660 &  &  \\
2109-04-28.0 & 1.26e-06 & 0.831 &  &  \\
2109-04-28.2 & 8.90e-08 & 1.240 &  &  \\
\hline
2110-05-2.9 & 4.12e-08 & 2.835 & 3.90e-08 & 2.864 \\
2110-05-2.5 & 2.61e-08 & 2.055 &  &  \\
2110-05-2.4 & 3.34e-08 & 1.925 &  &  \\
2110-05-2.4 & 2.89e-08 & 1.463 &  &  \\
2110-05-2.4 & 2.01e-09 & 0.190 &  &  \\
\hline
2112-04-28.0 & 1.93e-07 & 0.210 & 1.80e-07 & 0.214 \\
2112-04-27.9 & 3.18e-08 & 0.221 & 3.90e-10 & 0.149 \\
2112-04-28.4 & 5.23e-07 & 0.906 & 4.90e-07 & 0.906 \\
2112-04-28.3 & 2.49e-07 & 2.547 & 2.50e-07 & 2.573 \\
2112-04-28.4 & 1.55e-07 & 0.903 &  &  \\
\hline

2113-04-31.4 & 4.56e-07 & 0.115 & 4.50e-07 & 0.117 \\
2113-05-1.8 & 4.37e-08 & 0.027 & 9.30e-09 & 0.054 \\
2113-05-2.2 & 8.91e-10 & 3.718 &  &  \\
2113-05-1.8 & 1.68e-07 & 2.495 &  &  \\
2113-05-2.2 & 2.41e-08 & 0.005 &  &  \\
2113-05-1.9 & 7.93e-09 & 0.147 &  &  \\
\hline
2114-04-31.5 & 9.87e-09 & 0.084 & 9.50e-09 & 0.116 \\
2114-05-1.6 & 1.54e-09 & 2.235 &  &  \\
\hline
2115-04-28.5 & 3.32e-09 & 0.181 &  &  \\
2115-04-28.4 & 1.60e-08 & 0.980 &  &  \\
\hline
2116-04-30.1 & 1.39e-06 & 1.324 & 1.30e-06 & 1.338 \\
2116-04-30.4 & 2.92e-07 & 0.025 & 2.60e-07 & 0.035 \\
\hline
2117-05-2.2 & 1.85e-07 & 0.060 & 1.70e-07 & 0.061 \\
2117-04-31.3 & 6.86e-10 & 0.188 &  &  \\
\hline
2119-05-3.5 & 3.68e-08 & 0.234 & 3.50e-08 & 0.236 \\
2119-05-3.5 & 9.36e-07 & 0.385 & 8.20e-07 & 0.389 \\
2119-05-3.8 & 1.65e-10 & 0.106 &  &  \\
2119-05-3.6 & 1.05e-07 & 2.012 &  &  \\
\hline
2120-04-29.0 & 7.36e-10 & 2.812 &  &  \\
\hline
2121-04-28.2 & 3.90e-08 & 0.178 &  &  \\
2121-04-28.0 & 4.69e-09 & 0.192 &  &  \\
\hline
2122-05-3.6 & 2.11e-08 & 0.091 & 2.90e-07 & 0.238 \\
2122-05-3.1 & 2.18e-06 & 0.362 & 2.00e-06 & 0.366 \\
2122-05-3.5 & 5.13e-07 & 2.030 & 5.10e-07 & 2.051 \\
\hline
2123-04-30.7 & 9.69e-09 & 0.084 &  &  \\
\hline
2124-04-27.9 & 1.47e-09 & 0.163 &  &  \\
\hline
\end{supertabular}

%\caption{Comparison table for asteroid 2008 JL3}
\end{center}

\begin{table}[h]
\centering
\caption{Computation times and the number of VI searches for 5 considered asteroids.}
\label{tab:computation_times}
\begin{tabular}{lll}

Asteroid & $N_{VI}$ & $t_{comp}$, s\\
\hline
2008 EX5 & 965 &2738.8 \\
2005 QK76 & 555 &2627.7 \\
2023 DO & 2118 & 6216.5\\
2008 JL3 & 984 &2361.9\\
2000 SG344 & 8530 &15524.7\\
\hline
\end{tabular}

\medskip {'Asteroid' is the asteroid designation, $N_{VI}$ is the number of situations in which the virtual impactor (VI) search procedure was triggered, $t_{comp}$ is the total computation time, obtained in parallel mode using two threads on an Intel(R) Xeon(R) Gold 6230 CPU at 2.10\,GHz.}
\end{table}

\begin{figure*}
\centering
\includegraphics[width=0.38\textwidth]{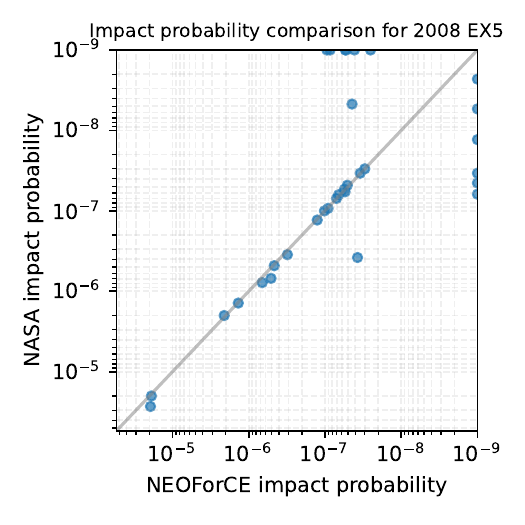} 
\includegraphics[width=0.38\textwidth]{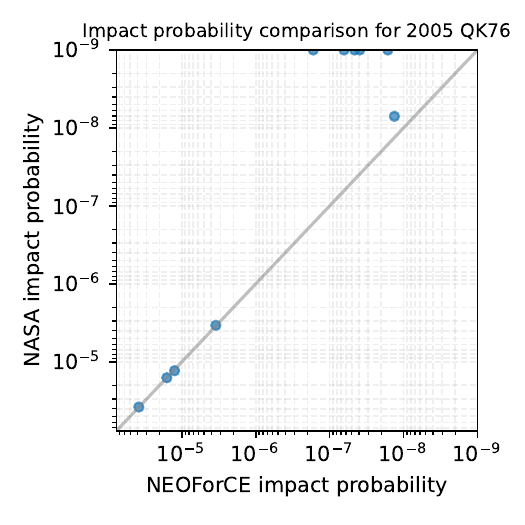} \\
\includegraphics[width=0.38\textwidth]{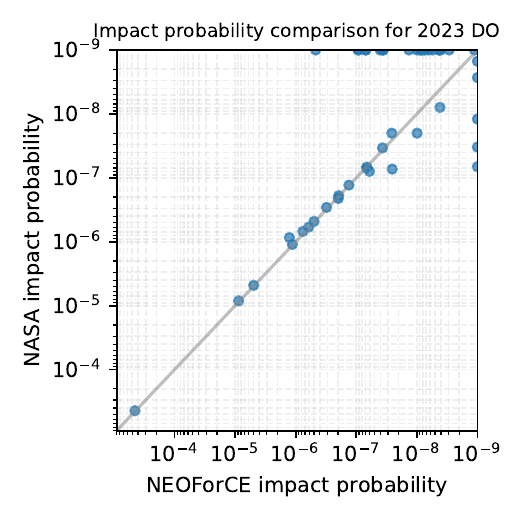} 
\includegraphics[width=0.38\textwidth]{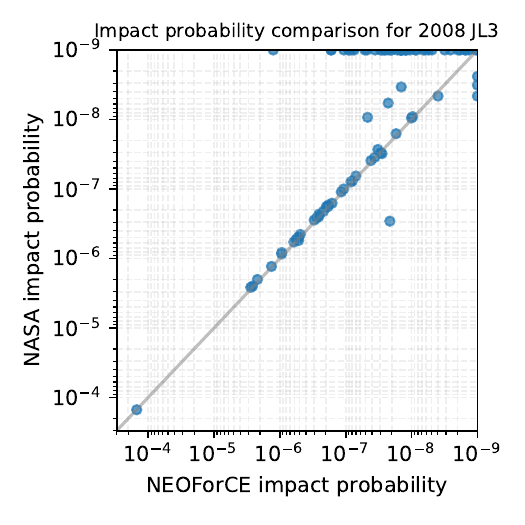}  \\
\includegraphics[width=0.38\textwidth]{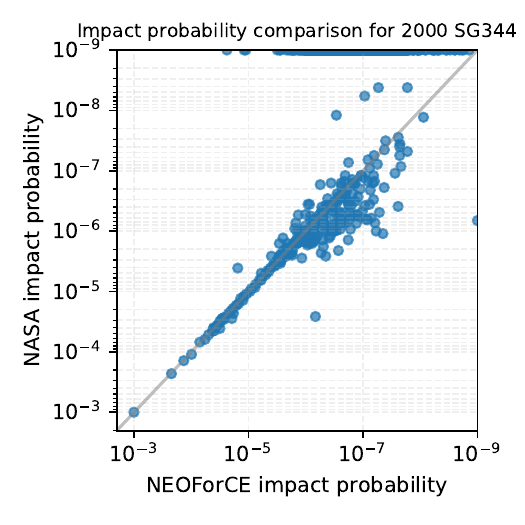} 
  \caption{  {Comparison between the NEOForCE impact monitoring system and Sentry-II.This figure summarizes the results from Tables 1–5. We show only impact probabilities larger than $10^{-9}$. A point located on the horizontal line at $P_{NASA} = 10^{-9}$ indicates a potential impact identified by NEOForCE but not detected by Sentry-II. Conversely, a point on the vertical line indicates an impact detected by Sentry-II but not found by NEOForCE.} }
     \label{fig:Comparison}
\end{figure*}

{A summary of the comparison for all five considered asteroids with NASA is shown in Fig.~\ref{fig:Comparison}. 
For asteroids 2008~EX5, 2005~QK76, 2023~DO, and 2008~JL3, we find good agreement between the two systems, with impact probability values consistent down to the $10^{-7}$ level, except for a few missing collision cases. 
For asteroid 2000~SG344, the agreement is noticeably worse and holds only down to about the $10^{-6}$ level. In this case, several potential collisions at the $10^{-5}$ probability level are missed by the NASA system. 
This behavior is likely caused by the very small relative velocity during close encounters, which reduces the completeness of the monitoring system (see Sect.~\ref{sec:completeness_problem}).
}

\section{Conclusions}

In this work we present NEOForCE (\textbf{N}ear-\textbf{E}arth \textbf{O}bject's \textbf{For}ecast of \textbf{C}ollisional \textbf{E}vents), a newly developed and fully independent monitoring system for predicting potential asteroid impacts with the Earth. The system is designed to operate autonomously from existing services and will ultimately use orbital solutions provided by the Paris Observatory. A central feature of NEOForCE is its search strategy: rather than testing encounters with the virtual asteroids (VAs) themselves, it identifies epochs when the Earth approaches the curvilinear uncertainty regions of the propagated VAs. This approach, combined with an assessment of the encounter's maximal impact probability, allows the system to efficiently narrow down possible collisions, and proceed to a targeted search for virtual impactors only when warranted. For the search of virtual impactors we use the semi-linear Partial Banana Mapping method, that can successfully find an impactor, even if the virtual asteroid is at 1\,au distance from the Earth.

We validated NEOForCE by comparing its performance with NASA’s Sentry-II system on five asteroids of different orbital and observational characteristics. The results demonstrate that NEOForCE reliably reproduces most of all significant impact solutions reported by Sentry-II above the $10^{-7}$ probability level, confirming the robustness of the method. Moreover, NEOForCE identified additional low-probability impact solutions (in the $10^{-7}$–$10^{-6}$ range) that Sentry-II did not report. These tests show that NEOForCE provides a reliable and complementary tool for impact monitoring, strengthening the independence and robustness of global planetary defense capabilities.

\begin{acknowledgements}
This project has received funding from the European Union’s Horizon 2020 research and innovation programme under the Marie Sklodowska-Curie grant agreement \#101068341 “NEOForCE”.
\end{acknowledgements}

%-------------------------------------------------------------------

\bibliographystyle{aa} % style aa.bst
\bibliography{bibliography} % your references Yourfile.bib

%\newpage

\end{document}